\documentclass[%
preprint,
nofootinbib,
 amsmath,amssymb,
 aps,
 11pt,
]{revtex4}

\usepackage{placeins}
\usepackage{multirow}
\usepackage{booktabs}
\usepackage{cancel}
\usepackage{color}
\usepackage{graphicx,physics,slashed}
\usepackage{graphicx}
\usepackage{hyperref}
\hypersetup{colorlinks=true, citecolor=blue, urlcolor=blue, linkcolor=blue}
\usepackage{dcolumn}
\usepackage{bm}
\usepackage{subfigure}
\usepackage{slashed}
\usepackage{setspace} 
\usepackage{ulem}
\usepackage{soul} 
\usepackage{cancel} 

\usepackage{float}
\usepackage{rotating}
\title{Lb->N1520}
\date{December 2024}

\begin{document}

\title{\Large Semileptonic Decay of $\Lambda_b \rightarrow N(1520)\ell^-\bar{\nu}_{\ell}$ from QCD Light-cone Sum Rules }
\vspace{5mm}
 
\author{
Ke-Sheng Huang$^{1}$\footnote{20251056@qhnu.edu.cn},
Ao-Sheng Xiong$^2$\footnote{xiongash2023@lzu.edu.cn},
Hua-Yu Jiang$^4$\footnote{jianghuayu@htu.edu.cn, corresponding author}, 
Fu-Sheng Yu$^{2,3}$\footnote{yufsh@lzu.edu.cn, corresponding author}
}

\address{$^1$College of Physics and Electronic Information Engineering, Qinghai Normal University, Xining 810000, China.\\
$^2$MOE Frontiers Science Center for Rare Isotopes and School of Nuclear Science and Technology, Lanzhou University, Lanzhou 730000, China. \\
$^3$Lanzhou Center for Theoretical Physics, Key Laboratory of Theoretical Physics of Gansu Province, Key Laboratory of Quantum Theory and Applications of MoE, Gansu Provincial Research Center for Basic Disciplines of Quantum Physics, Lanzhou University, Lanzhou 730000, China.\\
$^4$Institute of Particle and Nuclear Physics, Henan Normal University, Xinxiang 453007, China.}

\date{\today}
\vspace{5mm}
\begin{abstract}
 we investigate the complete set of vector and axial-vector form
 factors for the charged-current transition $\Lambda_b^0\to N(1520)^+$ using QCD light-cone sum rules (LCSRs) based on the light-cone distribution amplitudes (LCDAs) of the $\Lambda_b$ baryon, with the hard-scattering kernels evaluated at tree level. In the hadronic representation of the correlation function, we include the pole contributions of both the negative-parity $N(1520)$, with $J^P=3/2^-$, and the positive-parity $N(1720)$, with $J^P=3/2^+$. By matching a selected set of Lorentz structures, we derive sum rules that separate the $N(1520)$ contribution from the $N(1720)$ contribution within the adopted two-pole hadronic ansatz. After extrapolating the large-recoil LCSR results over the physical
$q^2$ region using a pole-improved $z$ expansion truncated at linear
order, we predict the differential branching fractions, the lepton forward-backward asymmetry, the charged-lepton polarization, and the $N(1520)$ polarization in $\Lambda_b^0\to N(1520)^+\ell^-\bar{\nu}_{\ell}$ ($\ell=e,\mu,\tau$). The corresponding total branching fractions are
$\mathcal{B}_{e,\mu,\tau}
=(12.4\pm10.5,\,12.4\pm10.5,\,5.0\pm4.3)\times10^{-5}$. Within the stated approximations, these results may serve as theoretical benchmarks for future experimental studies of this channel.

\end{abstract} 	


\maketitle
\vspace{10mm}

\section{Introduction}
\label{Introduction}
Exclusive semileptonic decays of beauty hadrons provide important tests of quark-flavor dynamics and nonperturbative QCD.  At the
quark level, charged-current $b\to u\ell\bar\nu_\ell$ transitions are governed
by the Cabibbo--Kobayashi--Maskawa (CKM) matrix element $V_{ub}$, whereas the
long-distance strong-interaction effects are encoded in hadronic form factors.
Improving the theoretical description of exclusive $b\to u$ processes remains
important because of the long-standing difference between the
inclusive and exclusive determinations of $|V_{ub}|$~\cite{ParticleDataGroup:2024cfk}. Bottom-baryon decays offer an independent route to this CKM coupling and give access to helicity observables beyond those available in pseudoscalar-to-pseudoscalar meson decays. The experimental potential of this program was demonstrated by LHCb through the measurement of $\Lambda_b^0\to p\mu^-\bar\nu_\mu$ relative to $\Lambda_b^0\to\Lambda_c^+\mu^-\bar\nu_\mu$~\cite{LHCb:2015eia}.  Combined with lattice QCD form factors~\cite{Detmold:2015aaa}, this measurement yielded a value of $|V_{ub}/V_{cb}|=0.083\pm0.004\,({\rm exp})\pm0.004\,({\rm lattice})$ and established baryonic semileptonic decays as competitive probes of quark-flavor physics.

Extending this program from the proton to excited nucleons probes a different aspect of strong-interaction dynamics.  The $N(1520)$ is a well-established isospin-$1/2$ resonance with spin and parity $J^P=3/2^-$~\cite{ParticleDataGroup:2024cfk}. The transition $\Lambda_b^0\to N(1520)^+\ell^-\bar\nu_\ell$ therefore probes a negative-parity spin-$3/2$ excitation of the light-quark system. Compared with the ground-state proton channel, the spin-$3/2$ final state leads to a richer set of helicity amplitudes and requires several independent vector and axial-vector form factors, even within the Standard Model. The differential rate is sensitive to the normalization and momentum dependence of the form factors, whereas the lepton forward-backward asymmetry and the polarizations of the charged lepton and the $N(1520)$ probe complementary combinations of helicity amplitudes. Such normalized observables do not depend on the overall $|V_{ub}|^{2}$ factor and are less sensitive to common form-factor normalization uncertainties.

Amplitude analyses of nonleptonic $\Lambda_{b}$ decays have demonstrated sensitivity to excited-nucleon contributions. LHCb observed the Cabibbo-suppressed decay $\Lambda_b^0\to J/\psi p\pi^-$ and found a rich structure in the $p\pi^-$ invariant-mass spectrum~\cite{LHCb:2014nhe}. A subsequent amplitude analysis explicitly included the neutral isospin partner $N(1520)^0$ and found its fitted contribution to be strongly dependent on the amplitude model~\cite{LHCb:2016lve}. These studies motivate a more detailed understanding of excited-nucleon amplitudes, but they neither constitute a measurement of the charged-current semileptonic mode nor constrain its form factors. No dedicated measurement of $\Lambda_b^0\to N(1520)^+\ell^-\bar\nu_\ell$ has yet been reported. Experimentally, the $N(1520)^+$
would be reconstructed through its strong-decay channels, such as $N\pi$, so comparison with data will ultimately require a finite-width amplitude analysis of
the corresponding $N\pi\ell^-\bar\nu_\ell$ final state. A calculation that treats the $N(1520)$ as an on-shell resonance therefore supplies transition form factors for the resonant contribution rather than a complete description of the measured multibody distribution.

The ground-state $\Lambda_b\to p$ form factors have been studied with lattice
QCD~\cite{Detmold:2015aaa} and with QCD light-cone sum rules (LCSRs)~\cite{Wang:2009hra,Khodjamirian:2011jp}.  In particular, LCSRs formulated in terms of the $\Lambda_b$-baryon light-cone distribution amplitudes (LCDAs) provide direct access to the form factors at large hadronic recoil.  The structure, modeling, and evolution
of the bottom-baryon LCDAs have been investigated systematically in heavy-quark
effective theory~\cite{Ball:2008fw,Ali:2012pn,Bell:2013tfa}. Perturbative LCSR analyses of the $\Lambda_b\to\Lambda$ and $\Lambda_b\to p$ transitions have also
shown that one-loop corrections, especially those associated with the
hard-collinear jet function, can be numerically important~\cite{Wang:2015ndk,Zhou:2026fui}. Missing radiative corrections must therefore be regarded as a separate source of uncertainty in a tree-level calculation.  For excited light baryons, LCSR studies of $\Lambda_{b,c}\to N^*(1535)$ and $\Lambda_b\to p,N^*(1535)$ have emphasized the sensitivity to interpolating currents, higher-twist effects, and the treatment of opposite-parity states \cite{Emmerich:2016jjm, Huang:2022lfr}.  The latter calculation is the closest existing LCSR precedent for a charged-current $\Lambda_{b}$ transition to an excited nucleon.

The spin-$3/2$ nature of the $N(1520)$ introduces further technical issues.
The electromagnetic transition $\gamma^*N\to N^*(1520)$ has been analyzed with
LCSRs, including the treatment of spin-$1/2$ contamination and opposite-parity spin-$3/2$ contributions~\cite{Aliev:2014eoa}. Helicity formalisms for $1/2\to3/2$ baryonic transitions have been developed in studies of the rare decay $\Lambda_b\to\Lambda(1520)\ell^+\ell^-$~\cite{Descotes-Genon:2019dbw, Das:2020cpv}.  For the same transition, form factors are available from lattice QCD \cite{Meinel:2020owd,Meinel:2021mdj}, dispersive analyses~\cite{Amhis:2022vcd}, the light-front quark model~\cite{Li:2022nim}, and a recent LCSR calculation based on $\Lambda_b$ LCDAs~\cite{Huang:2024oik}. Together, these studies provide useful methodological guidance. The studies of $\Lambda_b\to \Lambda(1520)$, however, concern the flavor-changing neutral-current $b\to s$ transition, and their form factors cannot be transferred to the charged-current $b\to u$ process. To the best of our knowledge, no dedicated LCSR calculation of the $\Lambda_b\to N(1520)$ charged-current form factors and
the associated semileptonic observables has been presented.

In this work, we calculate the four vector and four axial-vector form factors for the $\Lambda_b\to N(1520)$ transition using QCD LCSRs with the $\Lambda_b$-baryon LCDAs. The hard-scattering kernels are evaluated at tree level, and radiative corrections are not included. In the hadronic representation, we retain the negative-parity $N(1520)$ pole and the positive-parity $N(1720)$ pole. Within the adopted Dirac ordering, we select Lorentz structures that do not receive spin-$1/2$ contributions and solve the resulting linear system to separate the explicit $N(1720)$ pole contribution within this two-pole ansatz. Contributions from additional resonances, in particular the same-parity $N(1700)$, are not
resolved explicitly and constitute a limitation of the adopted hadronic representation. The LCSR calculation is assumed to be reliable in the large-recoil region. To study the full physical region, we use a pole-improved $z$-series expansion truncated after the term linear in $z$, without imposing additional relations at $q^2 = 0$ or at zero recoil. Predictions in the low-recoil region and
observables integrated over the full kinematic range therefore depend on this extrapolation, whose uncertainty is not determined by the LCSR calculation alone.
Subject to these approximations, we use the resulting form factors to study the differential rates and branching fractions for $\ell=e,\mu,\tau$, together with the lepton forward-backward asymmetry, the charged-lepton polarization, and the
$N(1520) $ polarization. The results should be interpreted as exploratory tree-level LCSR estimates for future studies of this channel rather than as precision Standard-Model predictions.

The remainder of this paper is organized as follows.  In Sec.~\ref{Framework}, we introduce the form-factor conventions and derive the light-cone sum rules.  The numerical inputs, sum-rule analysis, first-order $z$-series parameterization, and phenomenological observables are presented in Sec.~\ref{Results}. Our main conclusions are summarized in Sec.~\ref{summary}, and the coefficient functions in the partonic representation are collected in Appendix~\ref{sec:Appendix-A}.
\section{Light-cone sum rules for the $\Lambda_b\rightarrow N(1520)$ transition form factors}
\label{Framework}
\FloatBarrier
To distinguish the two spin-$\frac{3}{2}$ states of opposite parity retained
in the hadronic representation, we denote the negative-parity state
$N(1520)$ and the positive-parity state $N(1720)$ by $N_-^*$ and $N_+^*$,
respectively. The charged weak current is $J_\mu^{V-A}=J^{V}_{\mu}-J^{A}_{\mu}$, where $J^{V}_{\mu}=\bar{u}\gamma_\mu b$ and $J^{A}_{\mu}=\bar{u}\gamma_\mu\gamma_5 b$. Adopting the form-factor convention of Ref.~\cite{Gutsche:2017wag}, we parameterize the vector and axial-vector
current matrix elements for the $\Lambda_b(J^P=\frac{1}{2}^+)\to N_-^*(J^P=\frac{3}{2}^-)$ transition in terms of eight form factors as
\begin{align}
&\left\langle N_-^*(p,s')\left|\bar{u}\gamma_\mu b\right|
\Lambda_b(p+q,s)\right\rangle= \nonumber\\
&\bar{u}^{\alpha}(p,s')
\Bigg[
g_{\alpha\mu}f_{1-}^{V}(q^2)
+\frac{(p+q)_{\alpha}}{m_{\Lambda_b}}
\left(
f_{2-}^{V}(q^2)\gamma_\mu
+f_{3-}^{V}(q^2)\frac{p_\mu}{m_{\Lambda_b}}
+f_{4-}^{V}(q^2)\frac{q_\mu}{m_{\Lambda_b}}
\right)
\Bigg]u(p+q,s),
\nonumber\\
&\left\langle N_-^*(p,s')\left|\bar{u}\gamma_\mu\gamma_5 b\right|
\Lambda_b(p+q,s)\right\rangle=\nonumber\\
&\bar{u}^{\alpha}(p,s')
\Bigg[
g_{\alpha\mu}g_{1-}^{A}(q^2)
+\frac{(p+q)_{\alpha}}{m_{\Lambda_b}}
\left(
g_{2-}^{A}(q^2)\gamma_\mu
+g_{3-}^{A}(q^2)\frac{p_\mu}{m_{\Lambda_b}}
+g_{4-}^{A}(q^2)\frac{q_\mu}{m_{\Lambda_b}}
\right)
\Bigg]\gamma_5u(p+q,s),
\label{equ:FF}
\end{align}
Here, $\bar{u}^{\alpha}(p,s')$ is the adjoint Rarita--Schwinger spinor
of the outgoing $N_-^*$ baryon, whereas $u(p+q,s)$ is the Dirac spinor
of the incoming $\Lambda_b$ baryon. Their masses are denoted by
$m_{N_-^*}$ and $m_{\Lambda_b}$, their four-momenta by $p$ and $p+q$,
and their spin projections by $s'$ and $s$, respectively. The invariant form factors $f_{i-}^{V}(q^2)$ and $g_{i-}^{A}(q^2)$ ($i=1,\ldots,4$) are functions of the invariant momentum transfer $q^2$, where $q$ is the momentum transferred to the lepton pair.

To construct the LCSRs for the $\Lambda_b^0\to N(1520)^+$ transition form factors, we consider the following vacuum-to-\(\Lambda_b\) correlation function
\begin{equation}
\label{eq:correlator}
\Pi_{\alpha\mu}(p,q)=i\int d^4x\,e^{ip\cdot x}\left\langle 0\left|
\mathcal{T}\left\{j_\alpha(x),J_\mu^{V-A}(0)\right\}
\right|\Lambda_b(p+q,s)\right\rangle ,
\end{equation}
where $p$ is the momentum flowing through the nucleon-resonance
channel and $q$ is the momentum transferred to the lepton pair.
Following Ref.~\cite{Lee:2002jb}, we employ the proton-channel
vector-spinor interpolating current
\begin{equation}
\label{eq:N-current}
\begin{aligned}
j_\alpha(x)=\varepsilon_{abc}\Big[\big(u^{aT}(x)C\sigma_{\rho\lambda}d^b(x)\big)
 \sigma^{\rho\lambda}\gamma_\alpha u^c(x)-\big(u^{aT}(x)C\sigma_{\rho\lambda}u^b(x)\big)\sigma^{\rho\lambda}\gamma_\alpha d^c(x)
\Big],
\end{aligned}
\end{equation}
Here, the superscript $T$ denotes transposition in Dirac space,
$C$ is the charge-conjugation matrix,
$\sigma_{\rho\lambda}=i[\gamma_\rho,\gamma_\lambda]/2$, and
$a$, $b$, and $c$ are color indices.
The current $j_\alpha$ is neither parity selective nor a pure
spin-$\frac{3}{2}$ operator: it couples to spin-$\frac{3}{2}$
nucleon states of both parities and also to spin-$\frac{1}{2}$ states.
In the two-pole ansatz adopted below, the
$N^*_-$ and $N_+^*$ poles are retained explicitly. Within the parity convention of Eq.~(\ref{equ:FF}), the $N_+^*$ matrix elements are obtained by replacing
$f_{i-}^V\to f_{i+}^V$ and $g_{i-}^A\to g_{i+}^A$; the vector-current
matrix element contains a rightmost factor $\gamma_5$, whereas the
axial-vector-current matrix element does not. The spin-$\frac{1}{2}$ pole contributions have the generic form
\begin{equation}
\label{eq:1/2-pole}
\begin{aligned}
\left\langle0\left|j_\alpha\right|N_{1/2}(p)\right\rangle
=
\left(Ap_\alpha+B\gamma_\alpha\right)\Gamma\,u(p),
\end{aligned}
\end{equation}
where $\Gamma=1$ or $\gamma_5$, depending on the parity convention.
Following Refs.~\cite{Aliev:2023tpk,Huang:2024oik}, we therefore fix
the ordering of the Dirac matrices and retain only Lorentz structures
that do not receive such contributions. 
 
Within the LCSR framework, the correlation function in
Eq.~(\ref{eq:correlator}) is evaluated at both the hadronic
and partonic levels. The two representations are decomposed into the
same set of independent Lorentz structures, and the corresponding
invariant amplitudes are matched through dispersion relations. After
invoking quark--hadron duality to subtract the continuum contribution
above the effective threshold $s_0$ and performing a Borel transformation
with respect to $p^2$, one obtains a system of linear sum rules for the
transition form factors.  Within the adopted two-pole ansatz, solving this
system allows the explicit $N(1520)$ and $N(1720)$ pole contributions to
be separated. The Borel transformation also suppresses contributions
from higher excited states and removes subtraction terms.

At the hadronic level, the correlation function in
Eq.~(\ref{eq:correlator}) is represented by inserting a complete set of
intermediate states carrying the quantum numbers of the interpolating
current $j_\alpha$. For the Lorentz structures that have been explicitly
verified not to receive spin-$\frac{1}{2}$ pole contributions, the
spin-$\frac{3}{2}$ part of the correlation function can be written,
within the narrow-width two-pole ansatz, as
\begin{align}
\Pi_{\alpha\mu}^{\mathrm{had}}(p,q)
&={}
\frac{1}{m_{N_-^*}^{\,2}-p^2}
\sum_{s'}
\left\langle 0\left|j_\alpha(0)\right|N_-^*(p,s')\right\rangle
\left\langle N_-^*(p,s')\left|J_\mu^{V-A}(0)\right|
\Lambda_b(p+q,s)\right\rangle
\nonumber\\
&+
\frac{1}{m_{N_+^*}^{\,2}-p^2}
\sum_{s'}
\left\langle 0\left|j_\alpha(0)\right|N_+^*(p,s')\right\rangle
\left\langle N_+^*(p,s')\left|J_\mu^{V-A}(0)\right|
\Lambda_b(p+q,s)\right\rangle
+\cdots ,
\label{eq:hadronic-representation}
\end{align}
where the ellipsis represents all other resonance and continuum contributions.
The couplings of the interpolating current to the two explicitly
retained states define the corresponding decay constants
\begin{align}
\left\langle 0\left|j_\alpha(0)\right|N_+^*(p,s')\right\rangle
&=\lambda_+\,u_\alpha(p,s'),\nonumber\\
\left\langle 0\left|j_\alpha(0)\right|N_-^*(p,s')\right\rangle
&=\lambda_-\,\gamma_5u_\alpha(p,s'),
\label{eq:pole-residues}
\end{align}
Here, $\lambda_+$ and $\lambda_-$ are the decay constants associated with
the $N_+^*$ and $N_-^*$ states, respectively, and $u_\alpha(p,s')$ is
the corresponding Rarita--Schwinger spinor. In each matrix element, the
spinor is evaluated with the mass of the corresponding baryon. With the relations of the spin-$\frac{3}{2}$ baryon Rarita-Schwinger spinors and summing up the spin 
\begin{align}
\label{equ::RSsum}
\sum_{s'} u_\alpha(p,s')\bar{u}_\beta(p,s')=-\left(\slashed p+m_{N_\pm^*}\right)\Big[g_{\alpha\beta}
-\frac{1}{3}\gamma_\alpha\gamma_\beta
-\frac{2p_\alpha p_\beta}{3m_{N_\pm^*}^{\,2}}+\frac{p_\alpha\gamma_\beta-p_\beta\gamma_\alpha}
{3m_{N_\pm^*}}\Big]\ ,
\end{align}
the correlation function in the hadronic representation can be simplified. The possible spin-$\frac{1}{2}$ pole contributions originate from the coupling of the interpolating current $j_\alpha$ to spin-$\frac{1}{2}$ states and are proportional to $p_\alpha$ or $\gamma_\alpha$, which means the terms proportional to $p_\alpha$ or $\gamma_\alpha$ contain the contributions from the spin-$\frac{1}{2}$ states. Therefore, following Refs.~\cite{Aliev:2023tpk,Huang:2024oik}, we adopt the
fixed Dirac ordering displayed in Eq.~(\ref{eq:invariant-decomposition}) and discard invariant structures containing an explicit $p_\alpha$ or a leftmost
$\gamma_\alpha$, since such structures can receive spin-$\frac12$ pole contributions according to Eq.~(\ref{eq:1/2-pole}). We retain the $g_{\alpha\beta}$ component of the spin-$\frac32$ spin sum and use the resulting eight-structure basis below.

After applying the Lorentz-structure selection described above and using the equation of motion for the external $\Lambda_b$ spinor, the retained vector and axial-vector parts of the correlation function can each be decomposed into eight independent Lorentz--Dirac structures:
\begin{align}
\Pi_{\alpha\mu}^{X}(p,q)
&={}
\Gamma_X
\Bigg[
\Pi_1^{X}g_{\alpha\mu}
+\Pi_2^{X}g_{\alpha\mu}\slashed q
\nonumber\\
&\quad
+q_\alpha\Big(
\Pi_3^{X}p_\mu
+\Pi_4^{X}p_\mu\slashed q
+\Pi_5^{X}\gamma_\mu
+\Pi_6^{X}\gamma_\mu\slashed q
+\Pi_7^{X}q_\mu
+\Pi_8^{X}q_\mu\slashed q
\Big)
\Bigg]u_{\Lambda_b}(p+q,s) ,
\label{eq:invariant-decomposition}
\end{align}
where $X=V,A$, with $\Gamma_V=-\gamma_5$ and $\Gamma_A=1$. The invariant amplitudes $\Pi_i^{X}\equiv\Pi_i^{X}(p^2,q^2)$, with $i=1,\ldots,8$, are scalar
functions of the independent kinematic variables; their arguments are suppressed for brevity. Substituting the decay constants and transition matrix elements into the hadronic representation and performing the spin sum, the contributions
to the selected Lorentz--Dirac structures can be written as
\begin{align}
\left.\Pi_{\alpha\mu}^{\mathrm{had}}(p,q)\right|_{\mathrm{sel}}
={}&
-\gamma_5
\left[
\frac{\lambda_-}{m_-^2-p^2}
\mathcal H_{\alpha\mu}^{(+)}[f_-^V;m_-]
+
\frac{\lambda_+}{m_+^2-p^2}
\mathcal H_{\alpha\mu}^{(-)}[f_+^V;m_+]
\right]u_{\Lambda_b}(p+q,s)
\nonumber\\
&+
\left[
\frac{\lambda_-}{m_-^2-p^2}
\mathcal H_{\alpha\mu}^{(-)}[g_-^A;m_-]
+
\frac{\lambda_+}{m_+^2-p^2}
\mathcal H_{\alpha\mu}^{(+)}[g_+^A;m_+]
\right]u_{\Lambda_b}(p+q,s)
+\cdots ,
\label{eq:hadronic-selected}
\end{align}
where $M=m_{\Lambda_b}$, $m_\pm=m_{N_\pm^*}$, and $\mathcal H_{\alpha\mu}^{(\eta)}[h;m]$ is defined as below
\begin{align}
\mathcal H_{\alpha\mu}^{(\eta)}[h;m]
&={}
g_{\alpha\mu}h_1
\left(m+\eta M-\eta\slashed q\right)
+
q_\alpha\Bigg[
\frac{2Mh_2+(m+\eta M)h_3}{M^2}\,p_\mu
-\eta\frac{h_3}{M^2}\,p_\mu\slashed q
\nonumber\\
&+\frac{(\eta m-M)h_2}{M}\,\gamma_\mu
+\frac{h_2}{M}\,\gamma_\mu\slashed q
+\frac{(m+\eta M)h_4}{M^2}\,q_\mu
-\eta\frac{h_4}{M^2}\,q_\mu\slashed q
\Bigg],
\qquad \eta=\pm1 ,
\label{eq:hadronic-block}
\end{align}
Here, $f_\pm^V=(f_{1\pm}^V,f_{2\pm}^V,f_{3\pm}^V,f_{4\pm}^V)$ and $g_\pm^A=(g_{1\pm}^A,g_{2\pm}^A,g_{3\pm}^A,g_{4\pm}^A)$. For compactness, the common $q^2$ dependence of all form factors is suppressed. The ellipsis denotes additional resonance and continuum contributions. 

At the partonic level, the correlation function is evaluated by means
of a light-cone operator-product expansion around $x^2=0$, with $p^2$
taken sufficiently spacelike. At tree level, the $\bar u(0)$ field in
the weak current is contracted with each of the $u(x)$ fields in the
interpolating current. After summing the four Wick contractions, one
obtains
\begin{align}
\Pi_{\alpha\mu,\xi}^{(0)}(p,q)
={}&i\int d^4x\,e^{ip\cdot x}\,
\varepsilon_{abc}
(C\sigma_{\kappa\delta})_{\lambda\beta}
(\sigma^{\kappa\delta}\gamma_\alpha)_{\xi\gamma}
[\gamma_\mu(1-\gamma_5)]_{\rho\tau}
\nonumber\\
&\times\Big[
S_u^{(0)}(x)_{\gamma\rho}
\langle0|u_{\lambda}^{aT}(x)d_{\beta}^{b}(x)b_{\tau}^{c}(0)
|\Lambda_b(p+q,s)\rangle
\nonumber\\
&\hspace{1.2cm}
-3S_u^{(0)}(x)_{\beta\rho}
\langle0|u_{\lambda}^{aT}(x)d_{\gamma}^{b}(x)b_{\tau}^{c}(0)
|\Lambda_b(p+q,s)\rangle
\Big],
\label{eq:partonic-side}
\end{align}
Here, $\xi$ is the open Dirac index of the correlation function, and
$S_u^{(0)}(x)_{\eta\rho}\equiv\langle0|T\{u_\eta(x)\bar u_\rho(0)\}|0\rangle$.
The relative factor $-3$ follows from the sum of the four Wick contractions, together with $(C\sigma_{\kappa\delta})^T=C\sigma_{\kappa\delta}$ and the
equal-coordinate isospin relation $\Phi_{\lambda\beta\tau}(x)=-\Phi_{\beta\lambda\tau}(x)$. At leading power in the heavy-quark expansion, the full-QCD weak
current is matched onto its HQET counterpart. We adopt the convention
\begin{align}
&\varepsilon_{abc}
\langle0|u_{\lambda}^{aT}(t_1n)d_{\beta}^{b}(t_2n)
h_{v,\tau}^{c}(0)|\Lambda_b(v,s)\rangle
\nonumber\\
={}&\Big[
\frac18 f_{\Lambda_b}^{(2)}(\mu)\Psi_2(t_1,t_2;\mu)
 (\slashed{\bar n}\gamma_5C^{-1})_{\beta\lambda}
+\frac14 f_{\Lambda_b}^{(1)}(\mu)\Psi_3^s(t_1,t_2;\mu)
 (\gamma_5C^{-1})_{\beta\lambda}
\nonumber\\
&-\frac18 f_{\Lambda_b}^{(1)}(\mu)\Psi_3^\sigma(t_1,t_2;\mu)
 (i\sigma_{n\bar n}\gamma_5C^{-1})_{\beta\lambda}
+\frac18 f_{\Lambda_b}^{(2)}(\mu)\Psi_4(t_1,t_2;\mu)
 (\slashed n\gamma_5C^{-1})_{\beta\lambda}
\Big][u_{\Lambda_b}(v,s)]_\tau ,
\label{eq:Lambda-b-LCDAs}
\end{align}
Here, $p^\mu+q^{\mu}=m_{\Lambda_b}v^\mu$, $v^2=1$, and, on the light cone, $t=v\cdot x, \  n^\mu=\frac{x^\mu}{v\cdot x},\  \bar n^\mu=2v^\mu-n^\mu $, such that $n^2=\bar n^2=0$ and $n\cdot\bar n=2$. Straight Wilson lines connecting the nonlocal fields to the origin are understood. The LCDAs $\Psi_2$, $\Psi_3^s$, $\Psi_3^\sigma$, and $\Psi_4$ have twists $2$, $3$, $3$, and $4$, respectively. For the matrix elements in Eq.~(\ref{eq:partonic-side}), both light fields are located at $x^\mu=tn^\mu$, and therefore $t_1=t_2=t$. We follow the labeling convention of Refs.~\cite{Ball:2008fw,Bell:2013tfa}; Ref.~\cite{Ali:2012pn} interchanges the superscript labels of $f_{\Lambda_b}^{(1)}$ and $f_{\Lambda_b}^{(2)}$. At tree level, $\bar u\Gamma b=\bar u\Gamma h_v+ O(\alpha_s,\Lambda_{\rm QCD}/m_b)$; perturbative matching and power-suppressed corrections are not included.

For the subsequent LCSR calculation, it is convenient to introduce the momentum-space LCDAs through the Fourier transform
\begin{align}
\Psi_i(t_1,t_2;\mu)
={}&\int_0^\infty d\omega_1\int_0^\infty d\omega_2\,
e^{-i(\omega_1t_1+\omega_2t_2)}
\psi_i(\omega_1,\omega_2;\mu)
\nonumber\\
={}&\int_0^\infty \omega\,d\omega\int_0^1du\,
e^{-i\omega(ut_1+\bar u t_2)}
\psi_i(\omega,u;\mu),
\label{equ::Fourier}
\end{align}
where $ \omega=\omega_1+\omega_2,\ u=\omega_1/\omega,\ \bar u=1-u,$ and
$\psi_i(\omega,u;\mu)\equiv\psi_i(u\omega,\bar u\omega;\mu)$. The integration measure follows from $d\omega_1d\omega_2=\omega\,d\omega\,du$. Here, $\omega_1$ and $\omega_2$ are the light-cone momentum components carried by the $u$ and $d$ quarks, respectively. For the matrix elements entering the present correlation function,
both light-quark fields are located at $x^\mu=tn^\mu$. Hence,
$t_1=t_2=t=v\cdot x$, and Eq.~(\ref{equ::Fourier}) reduces to
\begin{align}
\Psi_i(t,t;\mu)
={}&\int_0^\infty\omega\,d\omega\,
e^{-i\omega t}\widehat{\psi}_i(\omega;\mu),
&
\widehat{\psi}_i(\omega;\mu)
\equiv\int_0^1du\,\psi_i(\omega,u;\mu),
\label{equ::Fourier2}
\end{align}
The theoretical properties and phenomenological models of the
three-particle $\Lambda_b$ LCDAs have been studied in
Refs.~\cite{Ball:2008fw,Ali:2012pn,Bell:2013tfa}.
For the numerical analysis, we adopt the one-parameter exponential
ansatz constructed in Ref.~\cite{Bell:2013tfa} in the
Wandzura--Wilczek approximation. At the input scale
$\mu_0=1\,\mathrm{GeV}$, the model functions are
\begin{align}
\psi_2(\omega,u;\mu_0)
&=\frac{\omega^2u\bar u}{\omega_0^4}
e^{-\omega/\omega_0},
\nonumber\\
\psi_3^s(\omega,u;\mu_0)
&=\frac{\omega}{2\omega_0^3}
e^{-\omega/\omega_0},
\nonumber\\
\psi_3^\sigma(\omega,u;\mu_0)
&=\frac{\omega(2u-1)}{2\omega_0^3}
e^{-\omega/\omega_0},
\nonumber\\
\psi_4(\omega,u;\mu_0)
&=\frac{1}{\omega_0^2}
e^{-\omega/\omega_0} ,
\label{equ::exp-model}
\end{align}
The same shape parameter is assumed for all four LCDAs. For a conservative symmetric variation, we take $\omega_0=(0.28\pm0.05)\,\mathrm{GeV},$ based on the Model-I result in Ref.~\cite{Wang:2015ndk}.

After inserting the momentum-space $\Lambda_b$-baryon LCDAs defined in
Eqs.~(\ref{eq:Lambda-b-LCDAs})--(\ref{equ::Fourier2}) into the partonic
correlation function in Eq.~(\ref{eq:partonic-side}), and carrying
out the Fourier integrations and Dirac algebra, we project the vector
and axial-vector parts of the correlation function onto the same
basis of eight independent Lorentz structures introduced in
Eq.~(\ref{eq:invariant-decomposition}). The corresponding invariant
amplitudes can be written as
\begin{align}
\Pi_i^{d,\mathrm{OPE}}(p^2,q^2)
={}&\sum_{n=1}^{2}
\int_0^\infty d\omega\int_0^1du\,
\frac{C_{i,n}^{d}(\omega,u,q^2)}
     {\left[D(\omega,p^2,q^2)\right]^n},
\label{EQ:invaramp}
\end{align}
with the denominator
\begin{align}
D(\omega,p^2,q^2)
={}&\bar{\sigma}p^2+\sigma q^2
-\sigma\bar{\sigma}m_{\Lambda_b}^2-m_u^2 ,
\end{align}
Here, $\sigma=\omega/m_{\Lambda_b}\ \text{and}\ \bar{\sigma}=1-\sigma$. The index $d=\mathrm{V},\mathrm{A}$ labels the vector and axial-vector
current components, respectively. The index $i$ labels the Lorentz structures in
Eq.~(\ref{eq:invariant-decomposition}), while $n$ denotes the power of the partonic propagator denominator.  The functions $C_{i,n}^{d}(\omega,u,q^2)$ are the corresponding partonic coefficient functions. The factor of $\omega$ arising from the Jacobian $d\omega_1\,d\omega_2=\omega\,d\omega\,du$ is understood to be absorbed into these coefficient functions through the integrated LCDAs defined in Appendix~\ref{sec:Appendix-A}.

The invariant amplitudes obtained from the hadronic and OPE
representations are matched through dispersion relations in $p^2$.
Keeping the $N_-^*$ and $N_+^*$ pole contributions explicitly, the
resulting system of linear equations can be solved to isolate the
$N_-^*$ contribution and eliminate the explicit $N_+^*$ pole within
the adopted two-pole ansatz. Quark-hadron duality is used to subtract
the continuum contribution above the effective threshold $s_0$,
whereas the Borel transformation with respect to $p^2$ suppresses
contributions from higher-mass hadronic states. 

We denote by $\widehat{\Pi}_i^d(M^2,s_0,q^2)$, with
$d=V,A$, the continuum-subtracted and Borel-transformed OPE invariant
amplitudes multiplying the Lorentz structures defined in
Eq.~(\ref{eq:invariant-decomposition}). In defining
$\widehat{\Pi}_i^V$, the overall factor $-\gamma_5$ in the vector
part of the correlation function is understood to be factored out.
For compactness, we introduce
\begin{align}
\mathcal{N}_-=
\frac{e^{m_-^2/M^2}}{\lambda_-(m_-+m_+)} ,
\end{align}
The resulting sum rules are
\begin{align}
\label{eq:vector-ff-lcsr}
f^V_{1-}(q^2)={}&0, \nonumber\\
f^V_{2-}(q^2)={}&
\mathcal{N}_-m_{\Lambda_b}
\left[\widehat{\Pi}_5^V+
(m_{\Lambda_b}+m_+)\widehat{\Pi}_6^V\right], \nonumber\\
f^V_{3-}(q^2)={}&
\mathcal{N}_-m_{\Lambda_b}^2
\left[\widehat{\Pi}_3^V+
(m_{\Lambda_b}-m_+)\widehat{\Pi}_4^V
-2\widehat{\Pi}_6^V\right], \nonumber\\
f^V_{4-}(q^2)={}&
\mathcal{N}_-m_{\Lambda_b}^2
\left[\widehat{\Pi}_7^V+
(m_{\Lambda_b}-m_+)\widehat{\Pi}_8^V\right],
\end{align}
and
\begin{align}
\label{eq:axial-ff-lcsr}
g^A_{1-}(q^2)={}&0, \nonumber\\
g^A_{2-}(q^2)={}&-
\mathcal{N}_-m_{\Lambda_b}
\left[\widehat{\Pi}_5^A+
(m_{\Lambda_b}-m_+)\widehat{\Pi}_6^A\right], \nonumber\\
g^A_{3-}(q^2)={}&
\mathcal{N}_-m_{\Lambda_b}^2
\left[\widehat{\Pi}_3^A+
(m_{\Lambda_b}+m_+)\widehat{\Pi}_4^A
-2\widehat{\Pi}_6^A\right], \nonumber\\
g^A_{4-}(q^2)={}&
\mathcal{N}_-m_{\Lambda_b}^2
\left[\widehat{\Pi}_7^A+
(m_{\Lambda_b}+m_+)\widehat{\Pi}_8^A\right] ,
\end{align}
Here, the arguments $(M^2,s_0,q^2)$ of
$\widehat{\Pi}_i^d$ have been suppressed for brevity. At tree level, and for the interpolating current and Lorentz projections
adopted in this work, the OPE invariant amplitudes multiplying
$g_{\alpha\mu}$ and $g_{\alpha\mu}\slashed q$ vanish,
$\widehat{\Pi}_{1,2}^{V,A}=0$. The corresponding two-pole matching
equations therefore give
\begin{equation}
 f_{1-}^{V}(q^{2})=g_{1-}^{A}(q^{2})=0 ,
\end{equation}
At the same order, the tree-level OPE coefficient functions satisfy,
term by term,
\begin{equation}
\begin{aligned}
 \Pi_{7}^{d,\text{OPE}}
 &=\Pi_{3}^{d,\text{OPE}}-2\Pi_{6}^{d,\text{OPE}},
 &\qquad
 \Pi_{8}^{d,\text{OPE}}
 &=\Pi_{4}^{d,\text{OPE}},
 && d=V,A,\\
 \Pi_{3}^{A,\text{OPE}}
 &=\Pi_{3}^{V,\text{OPE}},
 &
 \Pi_{4}^{A,\text{OPE}}
 &=-\Pi_{4}^{V,\text{OPE}},
 &\qquad
\Pi_{6}^{V,\text{OPE}}-\Pi_{6}^{A,\text{OPE}}
 &=m_{\Lambda_b}\Pi_{4}^{V,\text{OPE}}.
\end{aligned}
\end{equation}
These identities are preserved by the common Borel transformation and
continuum subtraction and hence imply
\begin{equation}
 f_{3-}^{V}(q^{2})=f_{4-}^{V}(q^{2})
 =g_{3-}^{A}(q^{2})=g_{4-}^{A}(q^{2}) ,
\end{equation}
Both results are specific to the present tree-level LCSR calculation and
to its applicable $q^{2}$ region. In particular, the latter equality,
which includes the finite-$m_u$ terms retained here, is a relation among
the matched hadronic form factors rather than an equality at the partonic
level. It is neither a model-independent all-order QCD identity nor a
consequence of SCET large-recoil symmetry, and may be modified by $\mathcal{O}(\alpha_s)$ corrections, higher-particle-number LCDAs, omitted higher-twist terms, and other power-suppressed contributions. Finally, eliminating the explicit positive-parity $N(1720)$ pole does not
remove possible contamination from the same-negative-parity spin-$3/2$
state $N(1700)$, which remains a separate systematic uncertainty of the
two-pole approximation.

In practice, the dispersion relation, Borel transformation, and quark--hadron duality are implemented through the following replacement rules:
\begin{align}
\int_0^\infty d\omega\int_0^1du\,
\frac{C_{i,1}^d(\sigma,u,q^2)}{D}
\;&\longrightarrow\;
-m_{\Lambda_b}\int_0^1du\int_0^{\sigma_0}
\frac{d\sigma}{\bar{\sigma}}\,
C_{i,1}^d(\sigma,u,q^2)
\exp\left[-\frac{s(\sigma,q^2)}{M^2}\right],
\nonumber
\\
\int_0^\infty d\omega\int_0^1du\,
\frac{C_{i,2}^d(\sigma,u,q^2)}{D^2}
\;&\longrightarrow\;
\frac{m_{\Lambda_b}}{M^2}
\int_0^1du\int_0^{\sigma_0}
\frac{d\sigma}{\bar{\sigma}^2}\,
C_{i,2}^d(\sigma,u,q^2)
\exp\left[-\frac{s(\sigma,q^2)}{M^2}\right]
\nonumber\\
&\quad+
\int_0^1du\,
\frac{\eta(\sigma_0)C_{i,2}^d(\sigma_0,u,q^2)}
     {m_{\Lambda_b}\bar{\sigma}_0^2}
\exp\left(-\frac{s_0}{M^2}\right) ,
\label{eq:Borel-rule-D2}
\end{align}
Here, $C_{i,n}^d(\sigma,u,q^2)$ is shorthand for
$C_{i,n}^d(\omega=m_{\Lambda_b}\sigma,u,q^2)$. The corresponding
coefficient functions are listed in
Appendix~\ref{sec:Appendix-A}. The functions entering the above
replacement rules are defined as
\begin{align}
s(\sigma,q^2)
&=\sigma m_{\Lambda_b}^2+
\frac{m_u^2-\sigma q^2}{\bar{\sigma}},
\ \ \ \eta(\sigma)=\left[
1+\frac{m_u^2-q^2}
        {\bar{\sigma}^2m_{\Lambda_b}^2}
\right]^{-1} ,
\label{eq:s-and-eta}
\end{align}
The upper integration limit $\sigma_0$ is the physical solution of
\begin{align}
s(\sigma_0,q^2)&=s_0,\qquad 0<\sigma_0<1,
\end{align}
namely,
\begin{align}
\sigma_0
={}&\frac{s_0+m_{\Lambda_b}^2-q^2
-\sqrt{\left(s_0+m_{\Lambda_b}^2-q^2\right)^2
-4m_{\Lambda_b}^2\left(s_0-m_u^2\right)}}
{2m_{\Lambda_b}^2},
&
\bar{\sigma}_0&=1-\sigma_0 .
\label{eq:sigma0}
\end{align}
Here, $s_0$ denotes the effective continuum threshold in the
$N^*$ channel.
\FloatBarrier

\section{Numerical analysis}
\label{Results}
\subsection{Numerical results for the form factors}
This subsection presents the numerical results for the
$\Lambda_b\to N(1520)$ form factors derived from the LCSRs in the
preceding section. In the next subsection, these form factors are used
to evaluate physical observables for the semileptonic decays
$\Lambda_b\to N(1520)\ell^-\bar{\nu}_{\ell}$, with
$\ell=e,\mu,\tau$. The numerical inputs required for the LCSR calculation
and the subsequent phenomenological analysis are summarized in
Table~\ref{Tab::input parameters}, while the LCSR-specific auxiliary
parameters are discussed separately below.
 \begin{table}
  \caption{Numerical values of the theoretical input parameters employed in LCSR prediction of the  $\Lambda_{b}\rightarrow N(1520)$ form factors as well as the subsequent phenomenological analysis for the $\Lambda_{b}\rightarrow N^{+}(1520)l^{-}\nu_{l^{-}}$ physical observables.}
  \label{Tab::input parameters}
  \centering
     \begin{tabular}{|l |l r||c |l r|}
     \hline
     \hline
     \text{Parameter}&\text{Value}&\text{Ref.}&\text{Parameter}&\text{Value}&\text{Ref.}\\
     \hline
        $m_{\Lambda_{b}}$ & 5.620\ \text{GeV}  &  \cite{ParticleDataGroup:2024cfk} &$|V_{ub}|$&$3.82\times 10^{-3}$&\cite{ParticleDataGroup:2024cfk}\\
        $m_{N^{*}_{-}}$& 1.510\ \text{GeV}  & \cite{ParticleDataGroup:2024cfk} &$G_{F}$&$1.166\times 10^{-5}$ GeV$^{-2} $&\cite{ParticleDataGroup:2024cfk} \\
        $m_{N^{*}_{+}}$& 1.680\ \text{GeV}  &   \cite{ParticleDataGroup:2024cfk}&$\tau_{\Lambda_{b}}$&$1.471\times 10^{-12}\ \text{s}$&\cite{ParticleDataGroup:2024cfk}\\
        $\lambda_{-}$& $0.127\pm 0.004$\ \text{GeV$^3$}& \cite{Azizi:2019dfh}&$m_{e}$&0.511 MeV&\cite{ParticleDataGroup:2024cfk}\\
        $m_{u}(\mu=1\ \text{GeV})$&$2.916\times 10^{-3}$\ \text{GeV}  &\cite{ParticleDataGroup:2024cfk}&$m_{\mu}$&0.106 GeV &\cite{ParticleDataGroup:2024cfk}   \\
        $f^{(1)}_{\Lambda_b}(\mu=1\ \text{GeV})$& $(0.030\pm 0.005)$\ \text{GeV$^3$}&\cite{Groote:1997yr} &$m_{\tau}$&1.777GeV&\cite{ParticleDataGroup:2024cfk}\\
        $f^{(2)}_{\Lambda_b}(\mu=1\ \text{GeV})$& $(0.030\pm 0.005)$\ \text{GeV$^3$} &\cite{Groote:1997yr} & $m_B$&5.279 GeV&\cite{ParticleDataGroup:2024cfk}\\
       &&&$m_{\pi}$&0.135 GeV&\cite{ParticleDataGroup:2024cfk}\\
       \hline
       \hline
     \end{tabular}    
 \end{table}

\begin{figure}
\centering
    \subfigure{\includegraphics[width=0.45\linewidth]{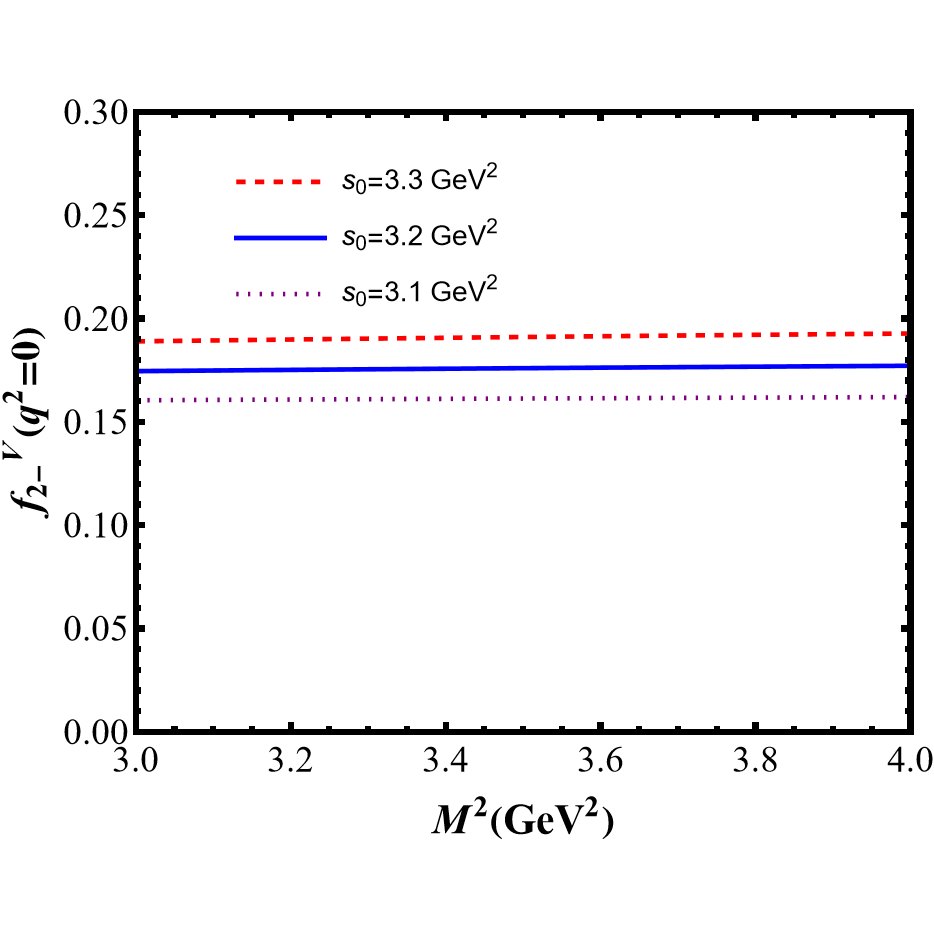}}
    \quad
     \subfigure{\includegraphics[width=0.45\linewidth]{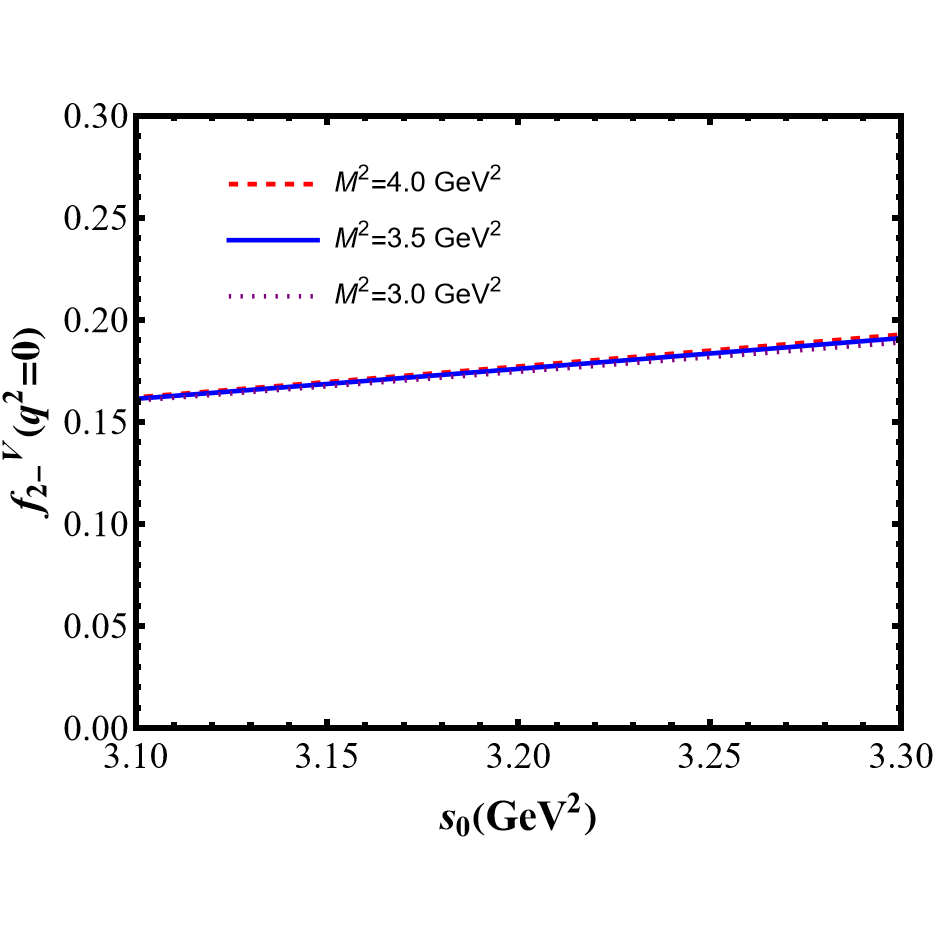}}
     \\
\caption{Dependence of $f_{2-}^{V}(0)$ on $M^2$ (left) and $s_0$
(right). In the left panel, the dashed, solid, and dotted curves
correspond to $s_0=3.3$, $3.2$, and $3.1~\mathrm{GeV}^2$, respectively.
In the right panel, they correspond to $M^2=4.0$, $3.5$, and
$3.0~\mathrm{GeV}^2$, respectively.}
\label{fig::dependence on M^2&s_0}
\end{figure}

The LCSRs involve two auxiliary parameters: the Borel parameter $M^2$ and the effective continuum threshold $s_0$. Although the physical form factors are independent of these parameters, the truncated LCSR predictions retain a residual dependence on them. The lower boundary of the Borel window is constrained by the
convergence of the twist expansion, whereas the upper boundary is constrained by the suppression of higher-state and continuum contributions. The effective continuum threshold $s_0$ specifies the onset of the unresolved higher-state and continuum contributions in the $N^*$ channel within the quark--hadron duality approximation. It is not a physical observable and cannot be determined uniquely. Its initial
range is guided by the hadronic spectrum in the $N^*$ channel and is chosen above the masses squared of the $N(1520)$ and $N(1720)$ poles retained explicitly in the hadronic representation. Its allowed range is then constrained together with $M^2$ by requiring an acceptable continuum contribution and a sufficiently stable LCSR
prediction. These considerations lead to the working intervals
\begin{equation}
M^2=(3.5\pm 0.5)\ \text{GeV}^{2},
\qquad
s_0=(3.2\pm 0.1)\ \text{GeV}^{2}  .
\label{eq:borel-window}
\end{equation}
In order to illustrate the residual dependence of the form factors on on $M^2$ and $s_0$, we show the dependence of the form factor $f^{V}_{2-}(0)$ on these two parameters as an example in Fig.~\ref{fig::dependence on M^2&s_0}. Similar dependencies are also observed for other $\Lambda_b \rightarrow N(1520)$ transition form factors. As displayed in Fig.~\ref{fig::dependence on M^2&s_0}, the dependence on $M^2$ is weak, and the variation with $s_0$ amounts to approximately $\pm9\%$ about the central value over the adopted interval.

The light-cone operator-product expansion of the correlation function
$\Pi_{\alpha\mu}(p,q)$ is reliable in the large-recoil region, corresponding
to low-to-intermediate values of $q^2$. We therefore use the LCSR predictions
in the fit window specified below to constrain a pole-improved linear $z$ parameterization, which is subsequently extrapolated to the full physical
region $m_\ell^2\leq q^2\leq q^2_{\rm max}$. The form factors are analytic in the
complex $q^2$ plane with a branch cut starting at $t_+$. The corresponding
conformal variable is defined as \cite{Bourrely:2008za,SentitemsuImsong:2014plu}
\begin{align}
\label{eq:z-variable}
z(q^2,t_0)=
\frac{\sqrt{t_+-q^2}-\sqrt{t_+-t_0}}
     {\sqrt{t_+-q^2}+\sqrt{t_+-t_0}} , 
\end{align}
This transformation maps the complex $q^2$ plane cut along
$q^2\geq t_+$ onto the unit disk, with the branch cut mapped onto
$|z|=1$. Following Ref.~\cite{Detmold:2015aaa}, we adopt the common
threshold $t_+=(m_B+m_\pi)^2$ corresponding to the lowest $B\pi$ continuum threshold in the crossed $b\bar u$ channel. We choose $t_0=t_-\equiv q^2_{\rm max}
=(m_{\Lambda_b}-m_{N^*_-})^2$, so that $z(q^2_{\rm max},t_0)=0$ at zero recoil.
After introducing a form-factor-dependent pole factor, we use the nominal
linear parameterization
\begin{align}
\label{eq:z-parameterization}
\mathcal{F}_i(q^2)=
\frac{\mathcal{F}_i(0)}
     {1-q^2/m_{\mathrm{pole},i}^{\,2}}
\left[
1+a_1^i\bigl(z(q^2,t_0)-z(0,t_0)\bigr)
\right] ,
\end{align}
where
$\mathcal{F}_i\in
\{f^V_{2-},f^V_{3-},f^V_{4-},g^A_{2-},g^A_{3-},g^A_{4-}\}$
denotes the six form factors that are nonzero at the order considered.
The pole masses $m_{\mathrm{pole},i}$ are listed in Table~\ref{Tab::pole mass}, while $\mathcal{F}_i(0)$ and $a_1^i$ are determined by fitting this parameterization to the LCSR predictions.
 \begin{table}
  \caption{Pole masses used in the $z$ parameterizations of the
$\Lambda_b\to N(1520)$ form factors. The values are taken from \cite{ParticleDataGroup:2024cfk,Detmold:2015aaa}.}
  \label{Tab::pole mass}
  \centering
     \begin{tabular}{|@{\hspace{0.7em}}c@{\hspace{0.7em}} |@{\hspace{0.7em}}c@{\hspace{0.7em}}||@{\hspace{0.7em}}c@{\hspace{0.7em}}|@{\hspace{0.7em}}c@{\hspace{0.7em}}|}
     \hline
     \hline
     \text{Form factors}&$m_{\text{pole}}$\text{(GeV)}& \text{Form factors}&$m_{\text{pole}}$\text{(GeV)}\\
     \hline
  $f^{V}_{2-},\ f^{V}_{3-} $&5.33& $g^{A}_{2-},\ g^{A}_{3-}$&5.71\\
  $f^{V}_{4-}$&5.66&$g^{A}_{4-}$&5.28\\
      \hline
      \hline
     \end{tabular}    
 \end{table}
\begin{figure}
\centering
    \subfigure{\includegraphics[width=0.31\linewidth]{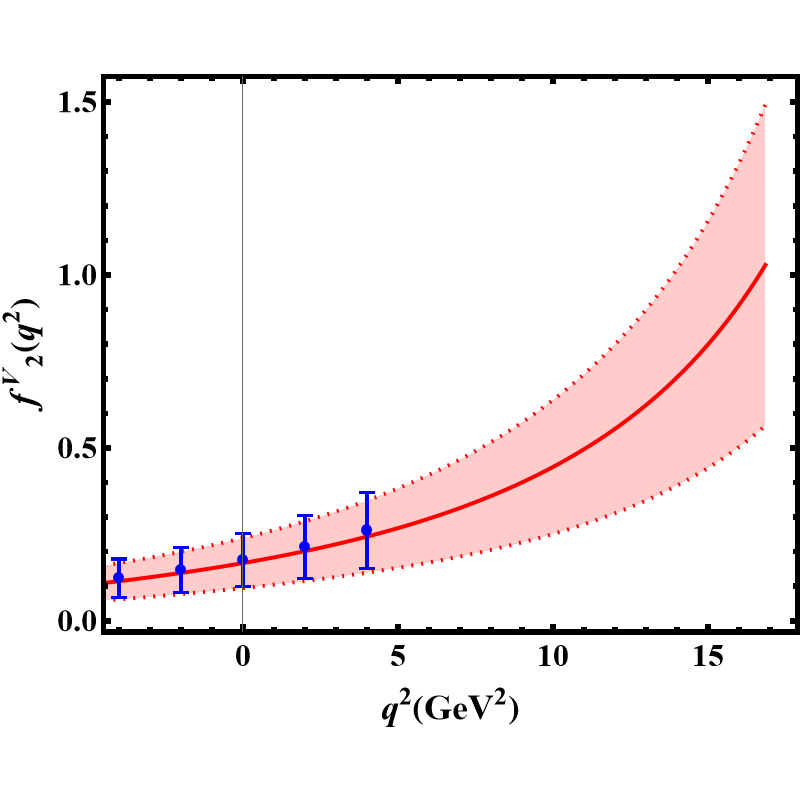}}
     \subfigure{\includegraphics[width=0.31\linewidth]{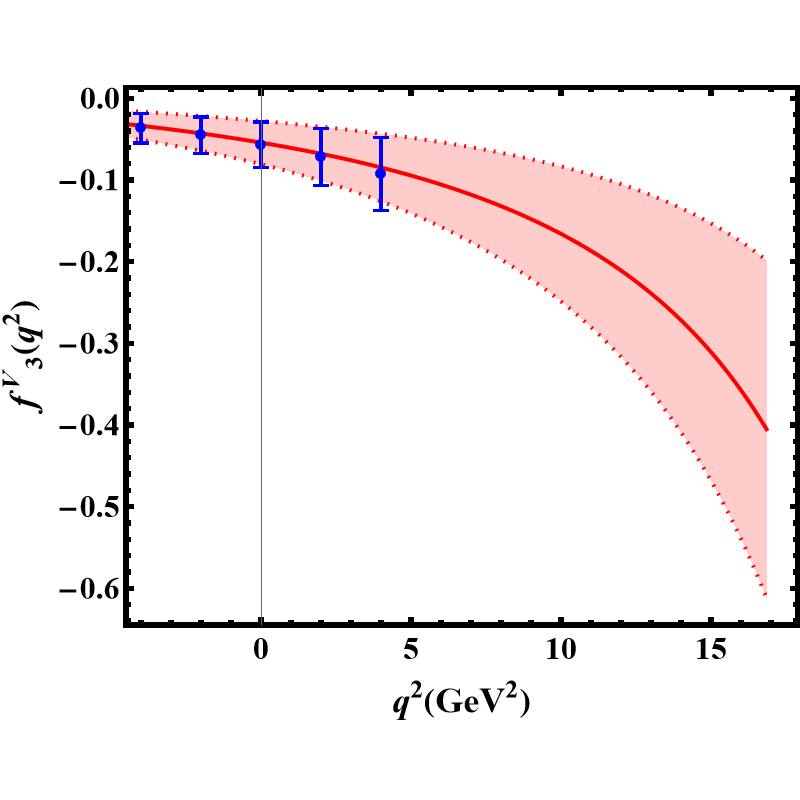}}
     \subfigure{\includegraphics[width=0.31\linewidth]{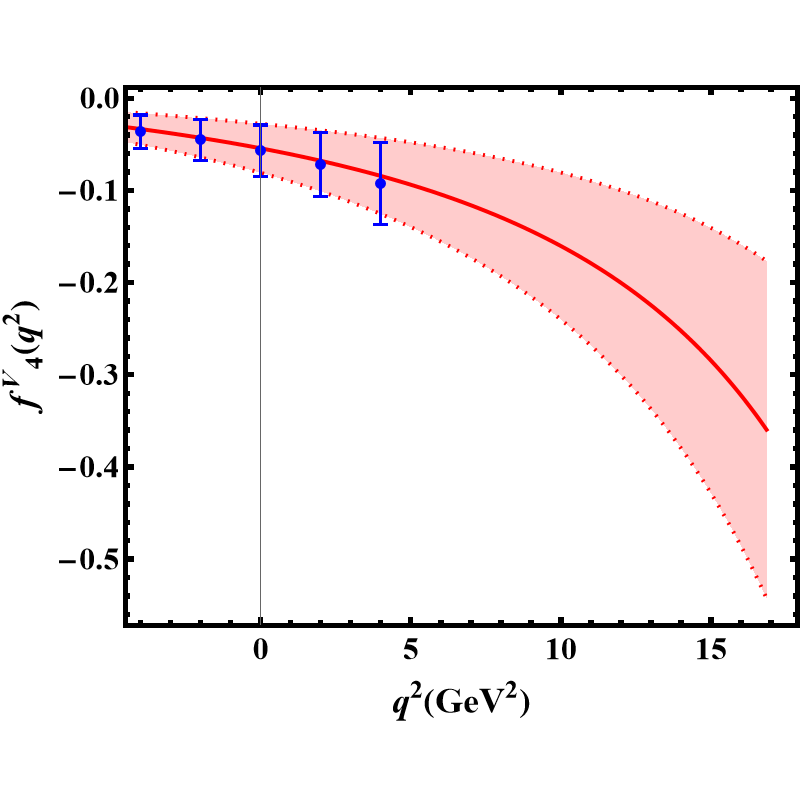}}
    \\
     \subfigure{\includegraphics[width=0.31\linewidth]{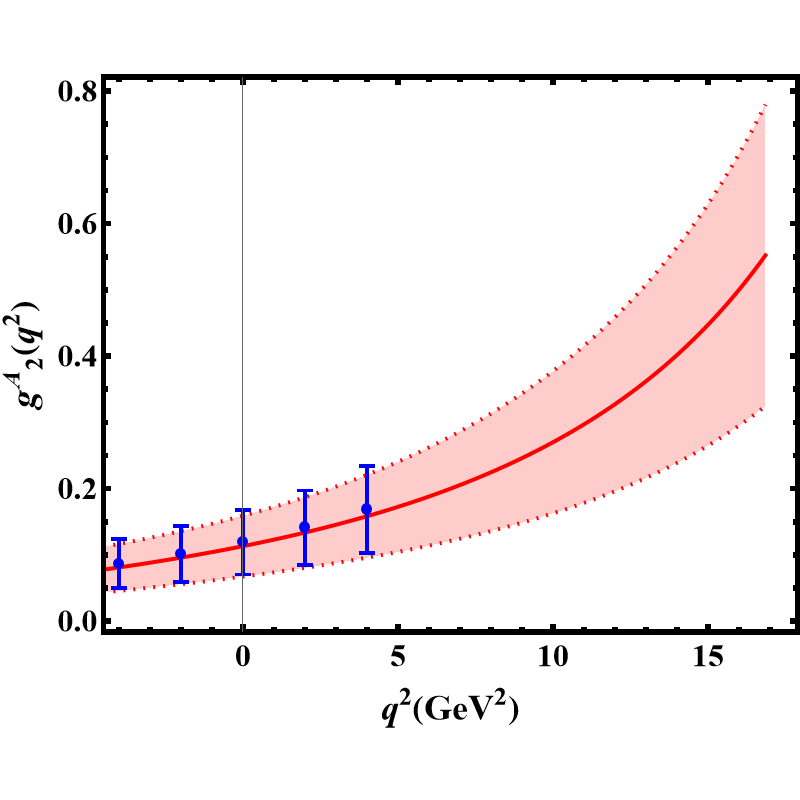}}
     \subfigure{\includegraphics[width=0.31\linewidth]{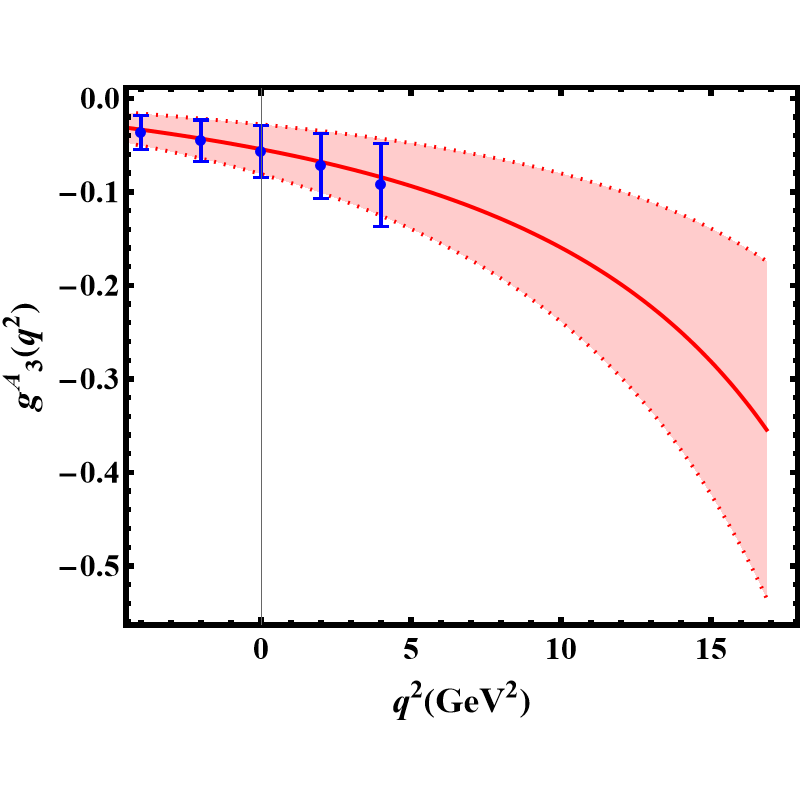}}
     \subfigure{\includegraphics[width=0.31\linewidth]{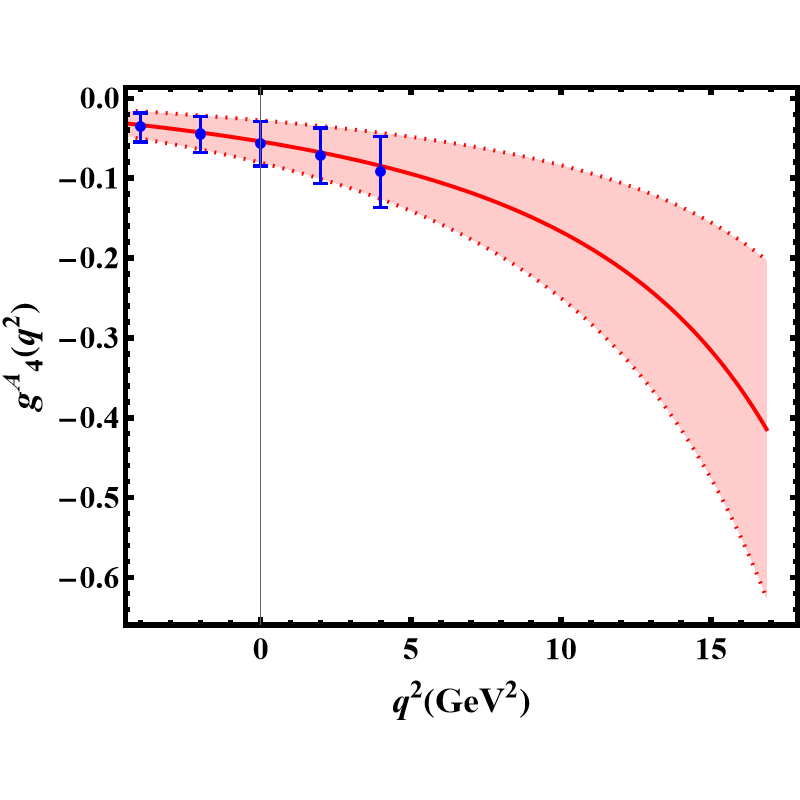}}
     \\
\caption{The $q^2$ dependence of the form factors for $\Lambda_{b}\rightarrow N(1520)$: where the red solid curves are the central values, the light red bands are the corresponding errors and the blue points are the samples used for the fitting at $q^2=\{-4.0,\ -2.0,\ 0.0,\ 2.0,\ 4.0\}$ GeV$^2$.}
\label{fig::dependence on q^2}
\end{figure}

To determine the  expansion coefficients, we perform a single $\chi^2$ fit to the LCSR predictions for the $\Lambda_b\rightarrow N(1520)$ form factors at five distinct kinematic points for each form factor:$q^2=\{-4.0,\ -2.0,\ 0,\ 2.0,\ 4.0\}\ \text{GeV}^2$. To do this fitting, we first generate $N=500$ ensembles of the input parameter set (including the Borel parameter $M^2$, the effective continuum threshold
$s_0$, the decay constants $\lambda_{-}$ and $f^{(1,2)}_{\Lambda_b}$ and the parameter $\omega_0$ of the $\Lambda_b$-LCDAs)  randomly \cite{SentitemsuImsong:2014plu} to construct the covariance matrix for this combined dataset. Since the data points are nearly linearly correlated, in order to incorporate more data points into the fitting process, we assume the presence of an additional small systematic error that reduces the off-diagonal elements of the covariance matrix to 92\% of their original values.  The corresponding loss function is built as a correlated $\chi^2$ function
\begin{equation}
\label{eq::the chi function}
\chi^2 = \left\{ \vec{F}_{\text{model}} - \vec{F}_{\text{data}} \right\}^T \cdot C_{\text{cov}}^{-1} \cdot \left\{ \vec{F}_{\text{model}} - \vec{F}_{\text{data}} \right\},
\end{equation}
where $\vec{F}_{\text{data}}$ denotes the vector of all input form-factor values, $\vec{F}_{\text{model}}$
is the corresponding vector of parameterized values, and $C_{\text{cov}}$ is the associated total covariance matrix. The $\chi^2$ is minimized to obtain the final fitted parameters.  The resulting parameters are summarized in
Table~\ref{Table:a^f_0}, and the fitted $q^2$ dependence is shown in
Fig.~\ref{fig::dependence on q^2}.
\begin{table}
    \caption{The fitted results of the $z$-series expansion coefficients, $\mathcal{F}_i(0)$ and $a_1^i$, for the form factors of the $\Lambda_b\rightarrow N(1520)$ transition.}
    \label{Table:a^f_0}
    \centering
    \begin{tabular}{|@{\hspace{0.7em}}c@{\hspace{0.7em}}|@{\hspace{0.7em}}c@{\hspace{1.5em}}c@{\hspace{0.7em}}|}
    \hline
    \hline
        Form factors & $\mathcal{F}_i(0)$ &  $a_1^i$\\
    \hline
        $f^{V}_{2-}(q^2)$&$0.167\pm0.072$&$-7.143\pm2.533$   \\
        $f^{V}_{3-}(q^2)$&$-0.054\pm0.026$&$-9.697\pm2.764$   \\
        $f^{V}_{4-}(q^2)$&$-0.054\pm0.027$&$-10.146\pm2.749$   \\
        $g^{A}_{2-}(q^2)$&$0.113\pm0.046$&$-6.440\pm2.429$   \\
        $g^{A}_{3-}(q^2)$&$-0.054\pm0.026$&$-10.207\pm2.746$   \\
        $g^{A}_{4-}(q^2)$&$-0.054\pm0.026$&$-9.621\pm2.767$   \\
        \hline
        \hline
    \end{tabular}
\end{table}

As shown in Table~\ref{Table:a^f_0} and Fig.~\ref{fig::dependence on q^2}, the fitted curves describe the LCSR results at the kinematic points included in the fit. The relative errors of the form factors are predominantly almost 50 \%, primarily due to their strong dependence on the uncertainty of the $\Lambda_{b}$-LCDAs parameter $\omega_{0}$. Therefore, more attention should be paid to the parameters of the $\Lambda_{b}$-LCDAs if we want to improve the precision of the form factors for the $\Lambda_{b}\rightarrow N(1520)$ transition.

\subsection{Semileptonic
$\Lambda_b\rightarrow N(1520)\ell^-\bar{\nu}_{\ell}$ decays}
\FloatBarrier
Using the form factors obtained in the preceding subsection, we present
predictions for the differential branching fraction
$d\mathcal{B}/dq^2$, the charged-lepton forward--backward asymmetry
$A_{FB}(q^2)$, the longitudinal polarization $P_B(q^2)$ of the
final-state $N(1520)$ baryon, and the longitudinal charged-lepton
polarization $P_{\ell}(q^2)$ for $\ell=e,\mu,\tau$. These observables
may serve as phenomenological benchmarks for future experimental
studies. In particular, extracting the polarization observables would
require sufficient signal yields and an angular analysis of the
$N(1520)$ decay products.

To construct these observables, we introduce the hadronic helicity
amplitudes. We work in the $\Lambda_b$ rest frame, with the $N_-^*$
baryon moving along the positive $z$ direction and the virtual
$W^{*-}$ moving along the negative $z$ direction. The helicity amplitudes are defined by
\begin{align}
\label{equ;;helicity amplitudes}
H^{V(A)}_{\lambda_{N^{*}_{-}}, \lambda_{W-}}=\epsilon^{\dag}_{\mu}(\lambda_{W^{-}})\langle N^{*}_{-}(p,\lambda_{N^{*}_{-}})|V^{\mu}(A^{\mu})|\Lambda_{b}(p+q,\lambda_{\Lambda_b})\rangle \ ,
\end{align}
where $\lambda_{\Lambda_{b}},\ \lambda_{N^{*}_{-}},\ \lambda_{W^{-}}$ denote the helicity of the $\Lambda_b$ baryon, the $N(1520)$ baryon and the off-shell $W^{-}$, respectively. Using the form-factor convention of Eq.~\eqref{equ:FF}, the independent
helicity amplitudes are~\cite{Gutsche:2017wag}
\begin{align}
\label{equ;;helicity amplitudes for v,A current}
H^{V}_{\frac{1}{2}t}&=\sqrt{\frac{2}{3}\frac{Q_{-}}{q^2}}\frac{Q_{+}}{2m_{\Lambda_b}m_{N^{*}_{-}}}\Big(f^{V}_{1-}(q^2)m_{\Lambda_b}+f^{V}_{2-}(q^2)M_{-}+f^{V}_{3-}(q^2)\frac{M_{+}M_{-}-q^2}{2m_{\Lambda_b}}+f^{V}_{4-}(q^2)\frac{q^2}{m_{\Lambda_{b}}}\Big)\ , \nonumber \\
H^{V}_{\frac{1}{2}0}&=\sqrt{\frac{2}{3}\frac{Q_{+}}{q^2}}\Big(f^{V}_{1-}(q^2)\frac{M_{+}M_{-}-q^2}{2m_{N^{*}_{-}}}+f^{V}_{2-}(q^2)\frac{Q_{-}M_{+}}{2m_{\Lambda_{b}}m_{N^{*}_{-}}}+f^{V}_{3-}(q^2)\frac{|\mathbf{p}|^2}{m_{N^{*}_{-}}}\Big)\ , \nonumber \\
H^{V}_{\frac{1}{2}1}&=\sqrt{\frac{Q_{+}}{3}}\Big(f^{V}_{1-}(q^2)-f^{V}_{2-}(q^2)\frac{Q_{-}}{m_{\Lambda_{b}}m_{N^{*}_{-}}}\Big)\ ,  \ \ \ \ \ \ \ H^{V}_{\frac{3}{2}1}=\sqrt{Q_{+}}f^{V}_{1-}(q^2)\ ,\nonumber \\
H^{A}_{\frac{1}{2}t}&=-\sqrt{\frac{2}{3}\frac{Q_{+}}{q^2}}\frac{Q_{-}}{2m_{\Lambda_b}m_{N^{*}_{-}}}\Big(g^{A}_{1-}(q^2)m_{\Lambda_b}-g^{A}_{2-}(q^2)M_{+}+g^{A}_{3-}(q^2)\frac{M_{+}M_{-}-q^2}{2m_{\Lambda_b}}+g^{A}_{4-}(q^2)\frac{q^2}{m_{\Lambda_{b}}}\Big)\ , \nonumber \\
H^{A}_{\frac{1}{2}0}&=-\sqrt{\frac{2}{3}\frac{Q_{-}}{q^2}}\Big(g^{A}_{1-}(q^2)\frac{M_{+}M_{-}-q^2}{2m_{N^{*}_{-}}}-g^{A}_{2-}(q^2)\frac{Q_{+}M_{-}}{2m_{\Lambda_{b}}m_{N^{*}_{-}}}+g^{A}_{3-}(q^2)\frac{|\mathbf{p}|^2}{m_{N^{*}_{-}}}\Big)\ , \nonumber \\
H^{A}_{\frac{1}{2}1}&=\sqrt{\frac{Q_{-}}{3}}\Big(g^{A}_{1-}(q^2)-g^{A}_{2-}(q^2)\frac{Q_{+}}{m_{\Lambda_{b}}m_{N^{*}_{-}}}\Big)\ ,  \ \ \ \ \ \ \ H^{A}_{\frac{3}{2}1}=-\sqrt{Q_{-}}g^{A}_{1-}(q^2)\ , 
\end{align}
where $Q_{\pm}$ is defined as $Q_{\pm}=(m_{\Lambda_b}\pm m_{N^{*}_{-}})^2-q^2$, $M_{\pm}=m_{\Lambda_{b}}\pm m_{N^{*}_{-}}$ and $|\mathbf{p}|$  is the three-momentum of the $N^{*}_{-}$ baryon, defined as $|\mathbf{p}|=\lambda^{1/2}(m^{2}_{\Lambda_{b}}, m^{2}_{N^{*}_{-}}, q^2)/(2m_{\Lambda_{b}})$. The negative helicity amplitudes can be obtained through the following relations
\begin{align}
\label{equ;;relation for HA}
H^{V}_{-\lambda_{N^*_{-}}, -\lambda_{W-}}=H^{V}_{\lambda_{N^*_{-}}, \lambda_{W-}},\ \ \ \ \  H^{A}_{-\lambda_{N^*_{-}}, -\lambda_{W-}}=-H^{A}_{\lambda_{N^*_{-}}, \lambda_{W-}}\ ,
\end{align}
and the total helicity amplitudes can be obtained by
\begin{align}
\label{equ;;total HA}
H_{\lambda_{N^*_{-}}, \lambda_{W-}}=H^{V}_{\lambda_{N^*_{-}}, \lambda_{W-}}-H^{A}_{\lambda_{N^*_{-}}, \lambda_{W-}}\ ,
\end{align}
The differential angular distribution for the semileptonic $\Lambda_b\rightarrow N(1520)\ell^{-}\bar{\nu}_{\ell^{-}}$ can be written as~\cite{Li:2022hcn} 
\begin{align}
\label{the differentical angular diatribution}
\frac{d^2\Gamma}{dq^2d\cos{\theta_{\ell}}}=|\frac{G_{F}}{\sqrt{2}}V_{ub}|^{2}\frac{\sqrt{Q_{+}Q_{-}}q^{2}(1-\hat{m}^{2}_{\ell})^2}{512\pi^{3}m^{3}_{\Lambda_{b}}}\times \big(A_{1}+A_{2}\cos{\theta_{\ell}}+A_{3}\cos{2\theta_{\ell}}\big)
\end{align}
where $G_{F}$ is the Fermi constant, $V_{ub}$ is the CKM matrix element, $\hat{m}^{2}_{\ell}=m^{2}_{\ell}/q^2$($m_{\ell}$ is the lepton mass with $\ell=e, \mu ,\tau$), $\theta_{\ell}$ is the angle between the positive direction of the three-momentum of the final baryon state $N(1520)$ and the lepton in the $q^{2}$ rest frame, and the angular coefficients $A_{1},\ A_{2}\  \text{and}\ A_{3} $ are given as:
\begin{align}
\label{eq:angular-coefficients}
A_1
={}&\frac{3+\widehat m_\ell^2}{2}
 \left(
 |H_{\frac32,+1}|^2+|H_{\frac12,+1}|^2
 +|H_{-\frac32,-1}|^2+|H_{-\frac12,-1}|^2
 \right) \notag\\
&+(1+\widehat m_\ell^2)
 \left(|H_{\frac12,0}|^2+|H_{-\frac12,0}|^2\right)
 +2\widehat m_\ell^2
 \left(|H_{\frac12,t}|^2+|H_{-\frac12,t}|^2\right),
 \notag\\
A_2
={}&2\left(
 |H_{-\frac32,-1}|^2+|H_{-\frac12,-1}|^2
 -|H_{\frac32,+1}|^2-|H_{\frac12,+1}|^2
 \right) \notag\\
&-4\widehat m_\ell^2\,
 \operatorname{Re}\left[
 H_{\frac12,0}^*H_{\frac12,t}
 +H_{-\frac12,0}^*H_{-\frac12,t}
 \right],
 \notag\\
A_3
={}&\frac{1-\widehat m_\ell^2}{2}
 \left(
 |H_{\frac32,+1}|^2+|H_{\frac12,+1}|^2
 +|H_{-\frac32,-1}|^2+|H_{-\frac12,-1}|^2
 \right) \notag\\
&-(1-\widehat m_\ell^2)
 \left(|H_{\frac12,0}|^2+|H_{-\frac12,0}|^2\right).
\end{align}
The differential decay width can be obtained by integrating out the $\cos{\theta_{\ell}}$
\begin{align}
\label{the diefferential decay width}
\frac{d\Gamma}{dq^2}=|\frac{G_{F}}{\sqrt{2}}V_{ub}|^{2}\frac{\sqrt{Q_{+}Q_{-}}q^{2}(1-\hat{m}^{2}_{\ell})^2}{512\pi^{3}m^{3}_{\Lambda_{b}}}\times (2A_{1}-\frac{2}{3}A_{3})\ ,
\end{align}
Additionally, other important physical observables, such as the leptonic forward-backward asymmetry $A_{FB}(q^2)$, the final hadron polarization $P_{B}(q^2)$, and the lepton polarization $P_{\ell}(q^2)$, are also expressed in terms of the helicity amplitudes. They can be defined as
\begin{align}
\label{equ;;physical observables}
A_{FB}(q^2)&=\frac{\int^{1}_{0}\frac{d\Gamma}{dq^2d\cos{\theta_{l}}}d\cos{\theta_{l}}-\int^{0}_{-1}\frac{d\Gamma}{dq^2d\cos{\theta_{l}}}d\cos{\theta_{l}}}{\int^{1}_{-1}\frac{d\Gamma}{dq^2d\cos{\theta_{l}}}d\cos{\theta_{l}}}=\frac{3A_{2}}{6A_{1}-2A_{3}}\ , \\
P_{B}(q^2)&=\frac{d\Gamma^{\lambda_{N^{*}_{-}}=(3/2,1/2)}/dq^2-d\Gamma^{\lambda_{N^{*}_{-}}=(-3/2,-1/2)}/dq^2}{d\Gamma/dq^2}\ ,\\
P_{\ell}(q^2)&=\frac{d\Gamma^{\lambda_{\ell}=1/2}/dq^2-d\Gamma^{\lambda_{\ell}=-1/2}/dq^2}{d\Gamma/dq^2}\ ,
\end{align}
respectively, where 
\begin{align}
\label{equ;;polarization HA}
\frac{d\Gamma^{\lambda_{N^{*}_{-}}=(3/2,1/2)}}{dq^2}&=\frac{4}{3}\Big(\big(2+\hat{m}^{2}_{\ell}\big)\big(|H_{1/2,0}|^{2}+|H^{2}_{1/2,1}|^{2}+|H_{3/2,1}|^{2}\big)+3\hat{m}^{2}_{\ell}|H_{1/2,t}|^{2}\Big)\ ,\\
\frac{d\Gamma^{\lambda_{N^{*}_{-}}=(-3/2,-1/2)}}{dq^2}&=\frac{4}{3}\Big(\big(2+\hat{m}^{2}_{\ell}\big)\big(|H_{-1/2,0}|^{2}+|H_{-1/2,-1}|^{2}+|H_{-3/2,-1}|^{2}\big)+3\hat{m}^{2}_{\ell}|H_{-1/2,t}|^{2}\Big)\ ,\\
\frac{d\Gamma^{\lambda_{\ell}=1/2}}{dq^2}&=\frac{4}{3}\hat{m}^{2}_{\ell}\Big(|H_{1/2,0}|^{2}+|H_{1/2,1}|^{2}+|H_{3/2,1}|^{2}+3|H_{1/2,t}|^{2}\nonumber \\
&\ \ \ \ \ \ \ \ \ \ \  +|H_{-1/2,0}|^{2}+|H_{-1/2,-1}|^{2}+|H_{-3/2,-1}|^{2}+3|H_{-1/2,t}|^{2}  \Big)\ ,\\
\frac{d\Gamma^{\lambda_{\ell}=-1/2}}{dq^2}&=\frac{8}{3}\Big(|H_{1/2,0}|^{2}+|H_{1/2,1}|^{2}+|H_{3/2,1}|^{2}+|H_{-1/2,0}|^{2}+|H_{-1/2,-1}|^{2}+|H_{-3/2,-1}|^{2}  \Big),
\end{align}
Here, we have factored out the common term
\begin{align}
\label{common term}
|\frac{G_{F}}{\sqrt{2}}V_{ub}|^{2}\frac{\sqrt{Q_{+}Q_{-}}q^{2}(1-\hat{m}^{2}_{\ell})^2}{512\pi^{3}m^{3}_{\Lambda_{b}}}\ .
\end{align}
For an observable $\mathcal O(q^2)=A_{FB}(q^2)$,
$P_{B}(q^2)$, or $P_\ell(q^2)$, its rate-weighted
average over a $q^2$ bin is defined by
\begin{align}
\label{eq:averaged-observable}
\langle\mathcal O\rangle_{[q_1^2,q_2^2]}
=
\frac{
 \displaystyle\int_{q_1^2}^{q_2^2}
 dq^2\,\mathcal O(q^2)\dfrac{d\Gamma}{dq^2}
}{
 \displaystyle\int_{q_1^2}^{q_2^2}
 dq^2\,\dfrac{d\Gamma}{dq^2}
}.
\end{align}
For the full physical region,
$q_1^2=m_\ell^2$ and
$q_2^2=(m_{\Lambda_b}-m_{N^*})^2$.
In deriving these expressions, the $N(1520)$ resonance is treated as
an on-shell final-state baryon. Finite-width effects and the subsequent
decay of the $N(1520)$ are not included.
 \begin{table}
  \caption{Predicted total branching fractions and rate-weighted averages
of $A_{FB}$, $P_B$, and $P_\ell$ for the semileptonic decays
$\Lambda_b\to N(1520)\ell^-\bar{\nu}_\ell$. The branching fractions are
given in units of $10^{-5}$, and the averages are evaluated over the full
physical $q^2$ range. }
  \label{Table:observables}
  \centering
     \begin{tabular}{|@{\hspace{0.7em}}c@{\hspace{0.7em}}|@{\hspace{0.7em}}c@{\hspace{0.7em}} @{\hspace{0.7em}}c@{\hspace{0.7em}} @{\hspace{0.7em}}c@{\hspace{0.7em}} @{\hspace{0.7em}}c@{\hspace{0.7em}}|}
     \hline
     \hline
     \text{Decay channel} &\text{Br}$(\times{10^{-5}})$&$\langle A_{FB}\rangle$&$\langle P_{B}\rangle$&$\langle P_{l}\rangle$\\
     \hline
      $\Lambda_{b}\rightarrow N(1520)e^{-}\bar{\nu_{e}}$ &$12.443\pm10.487$&$0.228\pm0.023$&$-0.967\pm0.073$&$\simeq-1.00$\\
       $\Lambda_{b}\rightarrow N(1520)\mu^{-}\bar{\nu_{\mu}}$&$12.407\pm10.457$&$0.223\pm0.023$&$-0.967\pm0.073$&$-0.985\pm0.002$\\
       $\Lambda_{b}\rightarrow N(1520)\tau^{-}\bar{\nu_{\tau}}$ &$5.009\pm4.265$&$0.002\pm0.022$&$-0.946\pm0.079$&$-0.352\pm0.056$\\
       \hline
       \hline
     \end{tabular}    
 \end{table}
\begin{figure}
\centering
\subfigure{\includegraphics[width=0.31\linewidth]{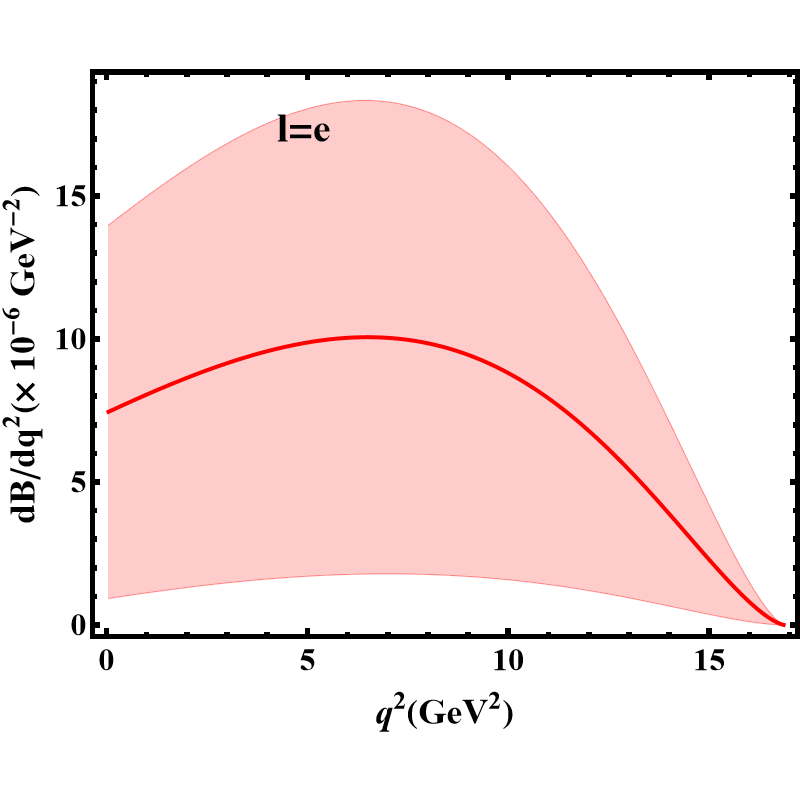}}
\subfigure{\includegraphics[width=0.31\linewidth]{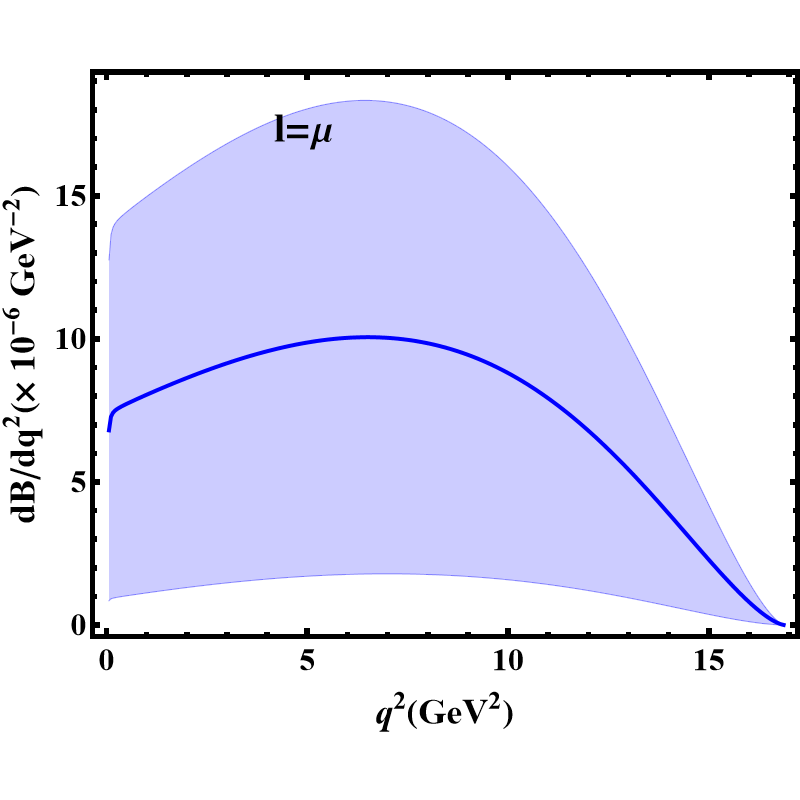}}
\subfigure{\includegraphics[width=0.31\linewidth]{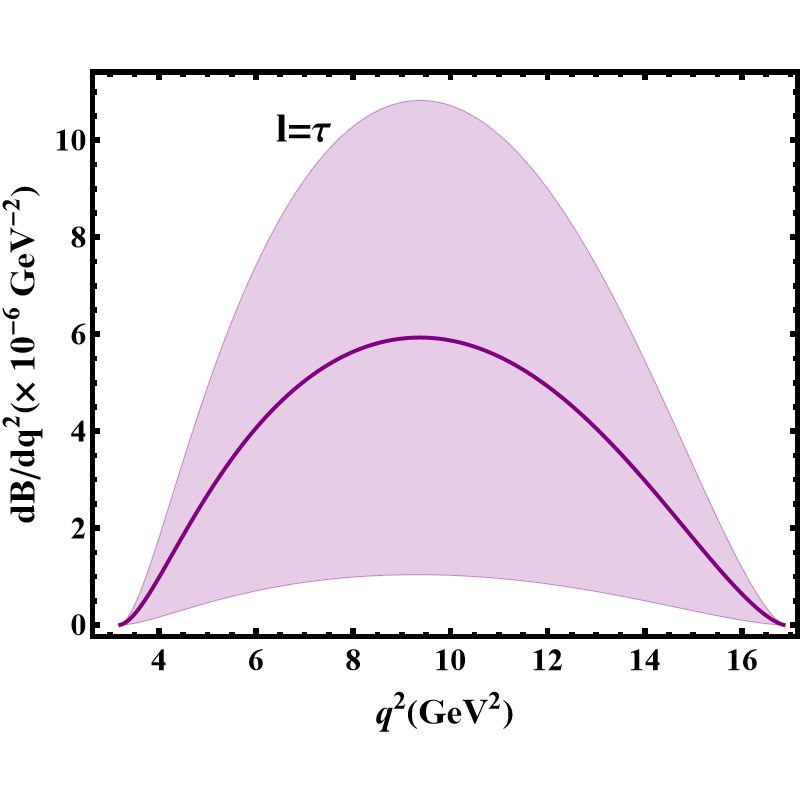}}
\\
\caption{The $q^2$ dependence of the differential branching fraction
$d\mathcal{B}/dq^2$ for $\Lambda_b\to N(1520)\ell^-\bar{\nu}_\ell$.
The left, middle, and right panels correspond to $\ell=e$, $\mu$, and
$\tau$, respectively. The solid curves denote the central predictions,
and the shaded bands represent the corresponding uncertainties.}
\label{fig::dependence on dBr}
\end{figure}
\begin{figure}
\centering
    \subfigure{\includegraphics[width=0.31\linewidth]{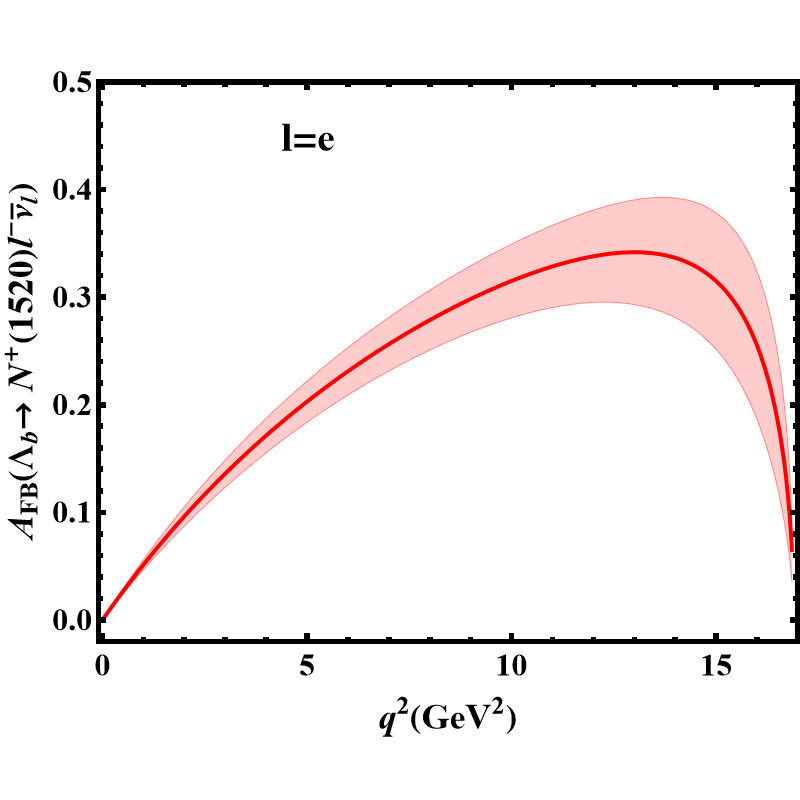}}
     \subfigure{\includegraphics[width=0.31\linewidth]{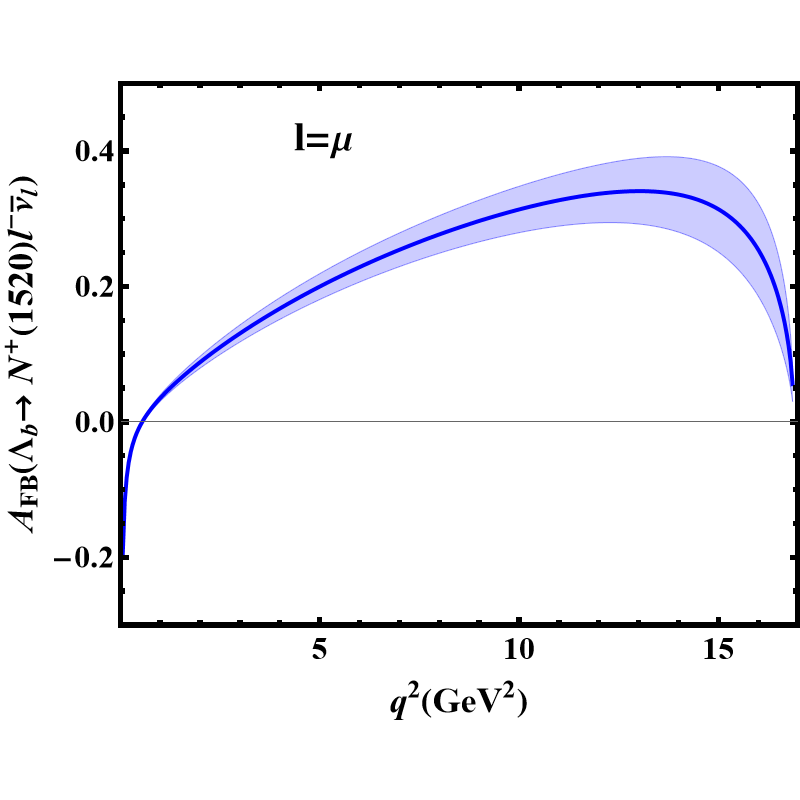}}
     \subfigure{\includegraphics[width=0.31\linewidth]{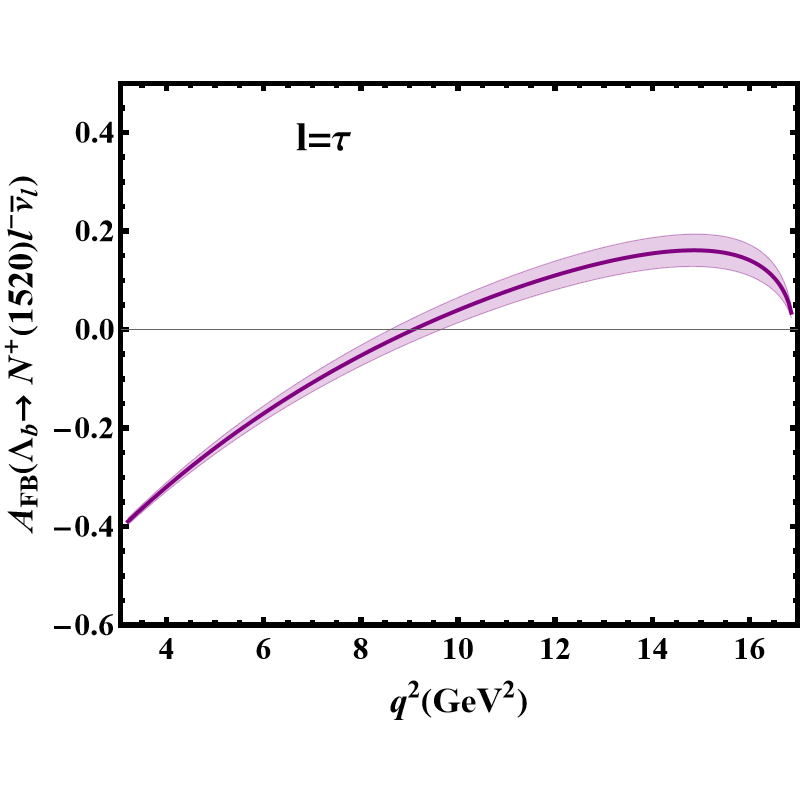}}
\caption{The $q^2$ dependence of the charged-lepton forward--backward asymmetry $A_{FB}(q^2)$ for $\Lambda_b\to N(1520)\ell^-\bar{\nu}_\ell$.
The left, middle, and right panels correspond to $\ell=e$, $\mu$, and
$\tau$, respectively. The solid curves denote the central predictions,
and the shaded bands represent the corresponding uncertainties.}
\label{fig::dependence on dAFB}
\end{figure}
\begin{figure}
\centering
\subfigure{\includegraphics[width=0.31\linewidth]{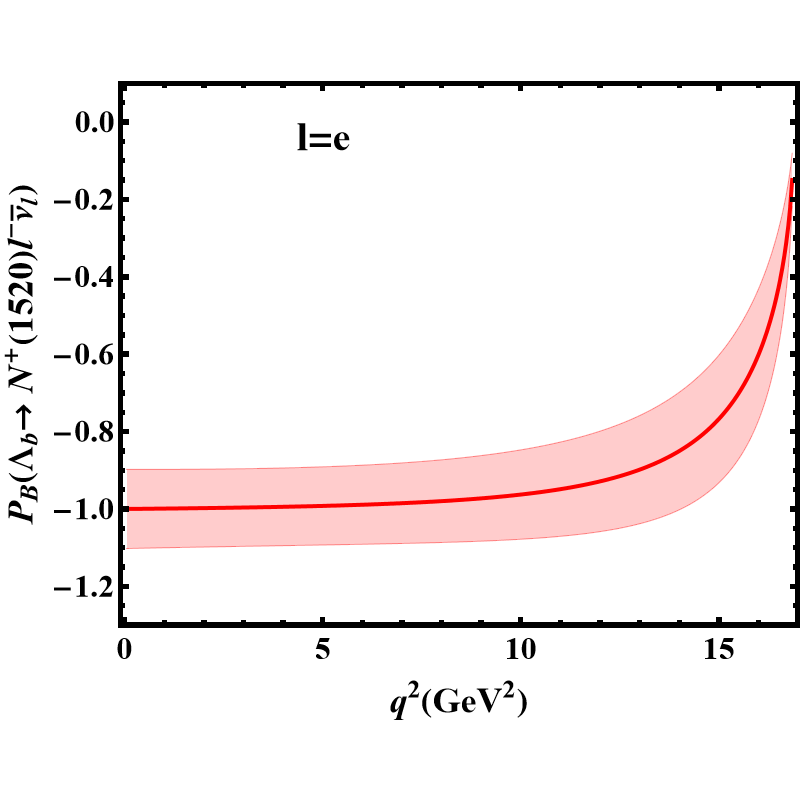}}
\subfigure{\includegraphics[width=0.31\linewidth]{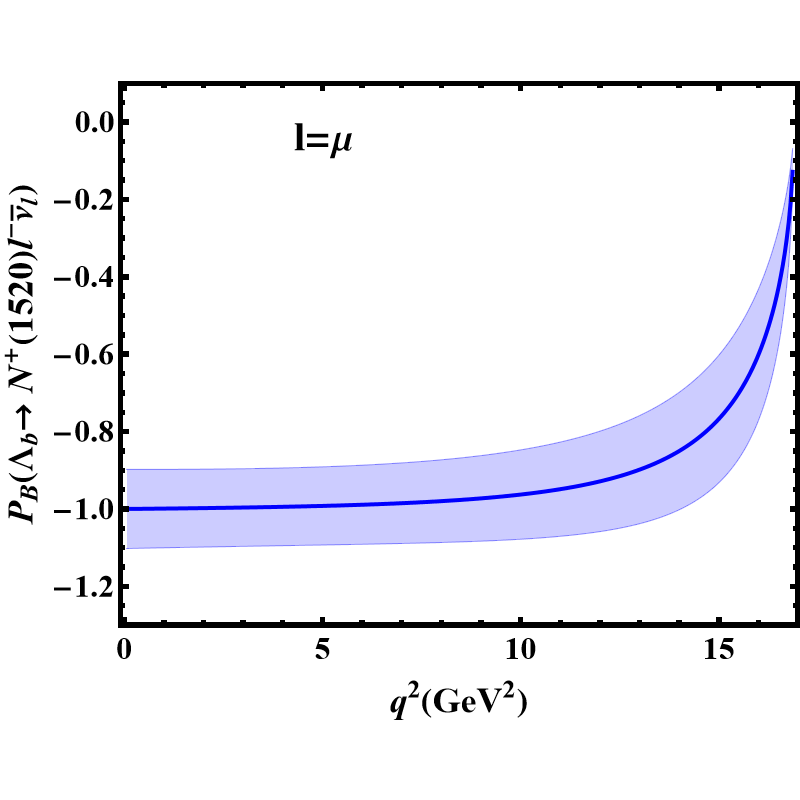}}
\subfigure{\includegraphics[width=0.31\linewidth]{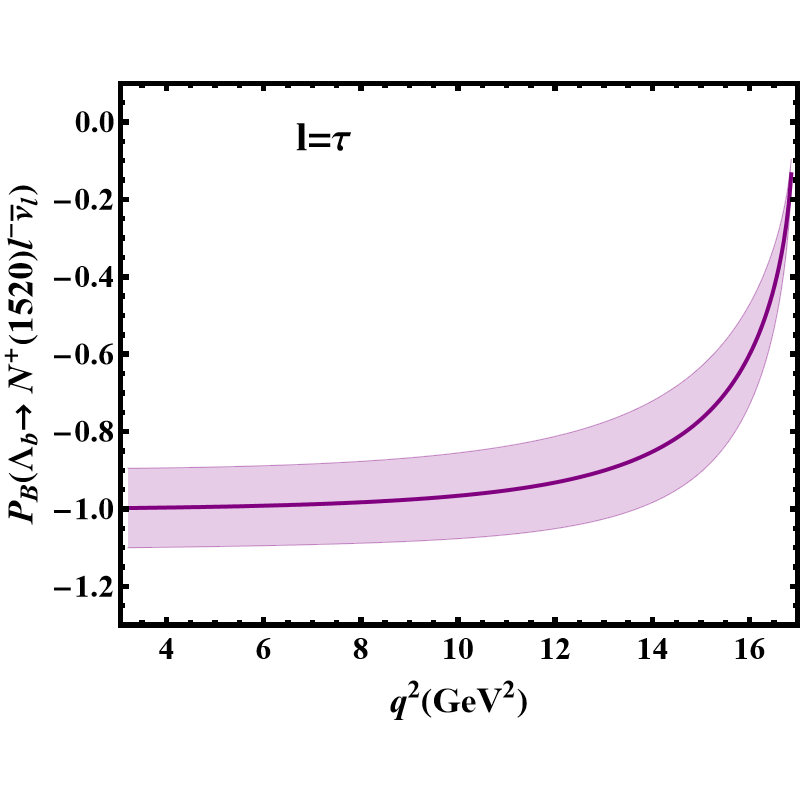}}
\caption{The $q^2$ dependence of the  final-state $N(1520)$ polarization $P_B(q^2)$ for $\Lambda_b\to N(1520)\ell^-\bar{\nu}_\ell$.
The left, middle, and right panels correspond to $\ell=e$, $\mu$, and
$\tau$, respectively. The solid curves denote the central predictions,
and the shaded bands represent the corresponding uncertainties.}
\label{fig::dependence on dPB}
\end{figure}
\begin{figure}
\centering
\subfigure{\includegraphics[width=0.31\linewidth]{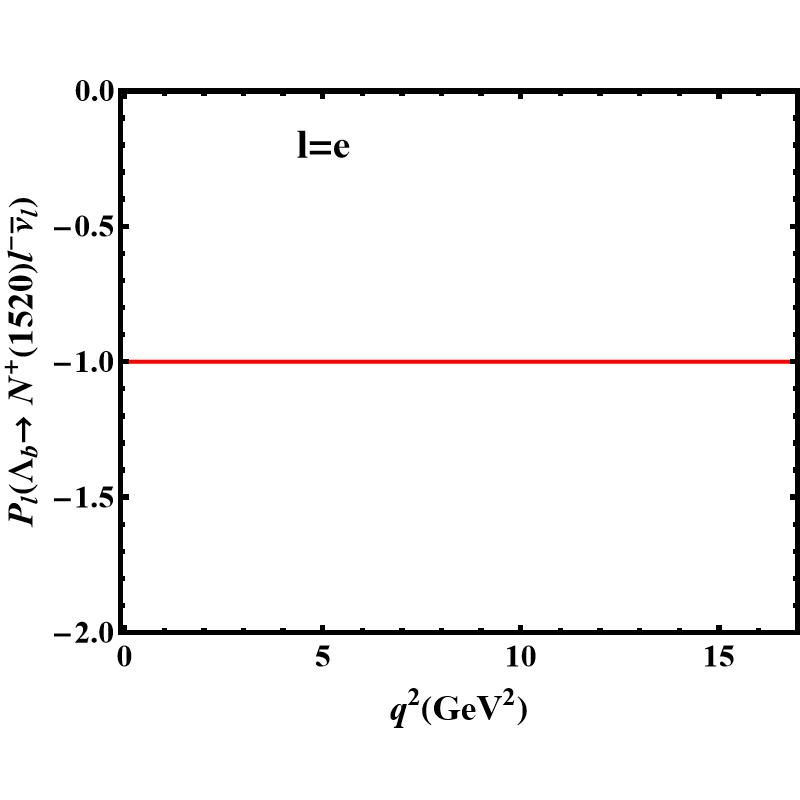}}
\subfigure{\includegraphics[width=0.31\linewidth]{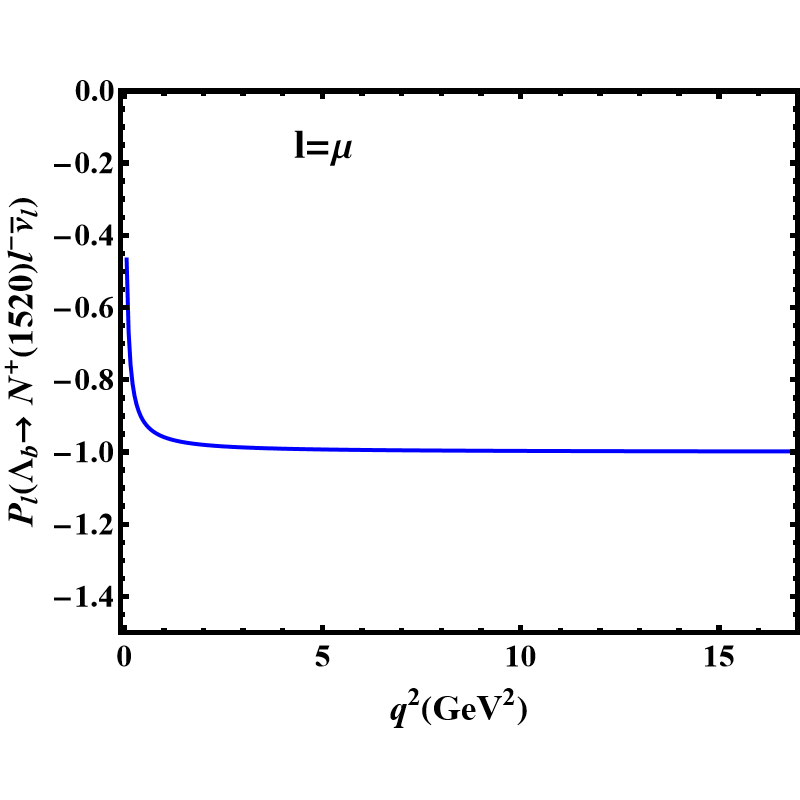}}
\subfigure{\includegraphics[width=0.31\linewidth]{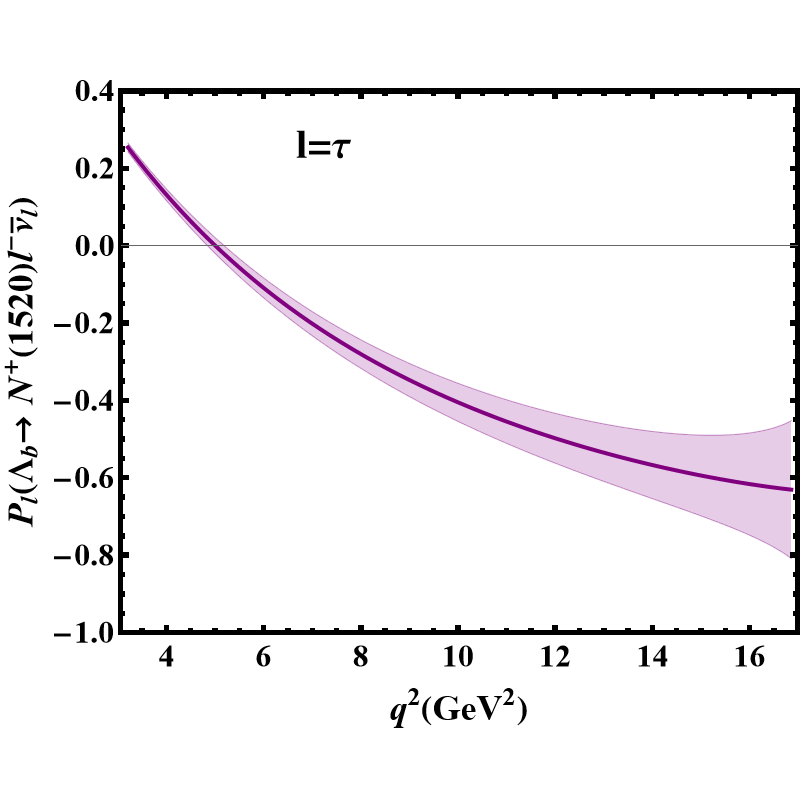}}
\caption{The $q^2$ dependence of the charged-lepton polarization $P_\ell(q^2)$ for $\Lambda_b\to N(1520)\ell^-\bar{\nu}_\ell$.
The left, middle, and right panels correspond to $\ell=e$, $\mu$, and
$\tau$, respectively. The solid curves denote the central predictions,
and the shaded bands represent the corresponding uncertainties.}
\label{fig::dependence on dPl}
\end{figure}

Using the fitted form factors, together with the CKM matrix element
$|V_{ub}|$ and the $\Lambda_b$-baryon lifetime $\tau_{\Lambda_b}$, we
obtain numerical predictions for the observables in the semileptonic
decays $\Lambda_b\to N(1520)\ell^-\bar{\nu}_\ell$. The total branching
fractions and the rate-weighted averages of $A_{FB}$, $P_B$, and $P_\ell$
are summarized in Table~\ref{Table:observables}, while their $q^2$
dependences are shown in Figs.~\ref{fig::dependence on dBr}--\ref{fig::dependence on dPl}. The quoted uncertainties are obtained by propagating the uncertainties of the fitted form factors.

The electron and muon modes have nearly identical branching fractions, as expected because their lepton-mass effects are small. By contrast, the
branching fraction of the tau mode is approximately $40\%$ of those of the light-lepton modes, owing to its reduced phase space and sizeable lepton-mass effects. The averaged forward--backward asymmetries in the electron and muon modes are positive and very similar. For the tau mode, $\langle A_{FB}\rangle=0.002\pm0.022$ is compatible with zero. As shown in Fig.~\ref{fig::dependence on dAFB}, this small averaged value results from a cancellation between the negative- and positive-$q^2$ regions across a
zero crossing, rather than from a uniformly small $A_{FB}(q^2)$. With the helicity convention adopted above, the large negative values of $\langle P_B\rangle$ indicate a strong dominance of the negative-helicity $N(1520)$ states. The electron and muon are predicted to be predominantly left-handed, as expected from the left-handed charged current. The less negative value of $\langle P_\ell\rangle$ in the tau mode arises from
finite-lepton-mass contributions to the positive-helicity amplitude. The branching fractions have relative uncertainties of approximately
$85\%$, inherited from the sizeable normalization uncertainties of the
LCSR form factors. In the adopted input-parameter analysis, the
sensitivity of the form factors to the $\Lambda_b$-LCDA parameter
$\omega_0$ makes an important contribution to these uncertainties.
The normalized asymmetries and helicity observables generally have
smaller absolute uncertainties because correlated normalization
uncertainties partially cancel in the corresponding ratios.


\FloatBarrier
\section{Summary}
\label{summary}
In this work, we have derived QCD light-cone sum rules for the four
vector and four axial-vector form factors governing the
$\Lambda_b^0\to N(1520)^+$ transition, using the LCDAs of the
$\Lambda_b$ baryon and evaluating the hard-scattering kernels at tree
level. At this order, $f^V_{1-}(q^2)=g^A_{1-}(q^2)=0$, while the remaining coefficient
functions yield $f^V_{3-}(q^2)=f^V_{4-}(q^2)=g^A_{3-}(q^2)=g^A_{4-}(q^2)$ within the present LCSR truncation. Using the fitted $q^2$ parameterizations of these form factors, we have studied the differential branching fractions, the charged-lepton forward--backward asymmetry $A_{FB}(q^2)$, the final-state $N(1520)$ polarization $P_B(q^2)$, and the charged-lepton polarization $P_\ell(q^2)$, together with their
rate-weighted averages, for the semileptonic decays $\Lambda_b^0\to N(1520)^+\ell^-\bar{\nu}_\ell$.

In the hadronic representation of the correlation function, we retain
the pole contributions of both the negative-parity $N(1520)$ state with
$J^P=3/2^-$ and the positive-parity $N(1720)$ state with $J^P=3/2^+$.
After fixing the ordering of the Dirac matrices, we select a set of eight
Lorentz structures that do not receive spin-$1/2$ pole contributions.
Matching the coefficients of these structures between the hadronic and
QCD representations yields a system of linear sum rules. Solving this
system separates the explicit $N(1520)$ and $N(1720)$ pole contributions
within the adopted two-pole hadronic ansatz. This procedure accounts for
the explicit $N(1720)$ contribution, but possible contributions from
additional resonances, including the same-parity $N(1700)$ state, are not
resolved and remain a systematic limitation of the analysis.

The form-factor uncertainties obtained in the adopted input-parameter
analysis are sizeable, with the $\Lambda_b$-LCDA parameter $\omega_0$
providing an important contribution. Improved knowledge of the
$\Lambda_b$ LCDAs is therefore required for more precise predictions.
The relations among the form factors quoted above are specific to the
present tree-level LCSR calculation and should not be interpreted as
model-independent QCD identities. Radiative corrections,
higher-particle-number LCDAs, omitted higher-twist terms, and other
power-suppressed contributions may modify these relations and are not
included in the quoted uncertainties.

Because the direct LCSR calculation is restricted to the large-recoil
region, we fit the LCSR predictions using a pole-improved $z$ expansion
truncated after the term linear in $z$ and use the resulting
parameterizations over the full physical $q^2$ range. Predictions in the
low-recoil region, and hence observables integrated over the full
kinematic range, therefore inherit a dependence on this extrapolation.
Using these form factors, we obtain $\mathcal{B}_{e,\ \mu,\ \tau}=\big(12.443\pm10.487,\,12.407\pm10.457,\,5.009\pm4.265\big)\times10^{-5}$ for the $e$, $\mu$, and $\tau$ modes, respectively. The electron and muon
modes have nearly identical branching fractions, whereas the tau mode is
suppressed by its reduced phase space and finite-lepton-mass effects.
The corresponding rate-weighted forward--backward asymmetries are
$0.228\pm0.023$, $0.223\pm0.023$, and $0.002\pm0.022$, respectively, so
that the tau-mode result is compatible with zero. We also predict a
strong dominance of negative-helicity $N(1520)$ states and predominantly
left-handed charged leptons in the light-lepton modes, while
finite-lepton-mass effects substantially reduce the magnitude of the tau
polarization. To the best of our knowledge, these results provide the
first set of theoretical benchmarks for future experimental studies of
the semileptonic decays
$\Lambda_b^0\to N(1520)^+\ell^-\bar{\nu}_\ell$.

\section{ACknowledgements}
This work is supported in part by the Natural Science Foundation of China under grant No. 12405114, No. 12335003, No. 12247101, and by the Fundamental Research Funds for the Central Universities under No. lzujbky-2023-stlt01, lzujbky-2024-oy02, lzujbky-2025-jdzx07 and lzujbky-2025-eyt01, the Scientific Research Innovation Capability Support Project for Young Faculty under Grant No. ZYGXQNJSKYCXNLZCXMP2, the Natural Science Foundation of Gansu Province (No.25JRRA799), and the ‘111 Center’ under Grant No. B20063.
\newpage
\appendix
\section{Partonic coefficient functions
$C_{i,n}^{d}(\sigma,u,q^2)$}
\label{sec:Appendix-A}
The coefficient functions entering Eq.~(\ref{EQ:invaramp}) for the
vector and axial-vector weak currents, $J_\mu^V=\bar u\gamma_\mu b$ and
$J_\mu^A=\bar u\gamma_\mu\gamma_5 b$, are listed below. Here $d=V,A$, $i=1,\ldots,8$, and $n=1,2$ denotes the power of the partonic denominator $D$.
\begin{itemize}
\item For the vector current $J^{V}_{\mu}=\bar{u}\gamma_{\mu}b$:
\begin{eqnarray}
C^{\rm V}_{1,n}&&=C^{\rm V}_{2,n}(n=1-2)=0\ , \ \ C^{V}_{3,1}=C^{V}_{4,1}=C^{V}_{8,1}=0\ ,\nonumber\\
C^{\rm V}_{3,2}&&=-\big[24\sigma\bar{\sigma}f^{(2)}_{\Lambda_b}(\overline{\psi_{4}}-\overline{\psi_{2}})+\frac{24(q^{2}-m^{2}_{\Lambda_{b}}\bar{\sigma}^{2}-m^{2}_{u})}{m^{2}_{\Lambda_{b}}}f^{(2)}_{\Lambda_b}\overline{\psi_{2}}+\frac{16m_{u}}{m_{\Lambda_{b}}}f^{(1)}_{\Lambda_b}\overline{\psi^{\sigma}_{3}}\big] \ ,\nonumber \\
C^{V}_{4,2}&&=C^{V}_{8,2}= -\big[\frac{16\sigma}{m_{\Lambda_{b}}}f^{(1)}_{\Lambda_b}\overline{\psi^{\sigma}_{3}}\big]\ ,\nonumber \\
C^{V}_{5,2}&&=-\big[12\sigma (m_{u}-\bar{\sigma}m_{\Lambda_{b}})f^{(2)}_{\Lambda_b}(\overline{\psi_{4}}-\overline{\psi_{2}})+\frac{8(m^{2}_{u}-\sigma q^{2}-m_{u}m_{\Lambda_{b}}\bar{\sigma})}{m_{\Lambda_{b}}\bar{\sigma}}f^{(1)}_{\Lambda_b}\overline{\psi^{\sigma}_{3}}\nonumber\\
&&\ \ \ +\frac{12(q^{2}-m^{2}_{\Lambda_{b}}\bar{\sigma}^{2}-m^{2}_{u})(m_{u}-\bar{\sigma}m_{\Lambda_{b}})}{m^{2}_{\Lambda_{b}}\bar{\sigma}}f^{(2)}_{\Lambda_b}\overline{\psi_{2}} \big]\ ,\nonumber \\
C^{V}_{5,1}&&=\big[\frac{12m_{u}}{m^{2}_{\Lambda_{b}}\bar{\sigma}}f^{(2)}_{\Lambda_b}\overline{\psi_{2}}-\frac{8\sigma}{m_{\Lambda_{b}}\bar{\sigma}}f^{(1)}_{\Lambda_b}\overline{\psi^{\sigma}_{3}}\big]\ ,  \nonumber \\
C^{V}_{6,2}&&=-\big[12\sigma f^{(2)}_{\Lambda_b}(\overline{\psi_{4}}-\overline{\psi_{2}})+\frac{12(q^{2}-m^{2}_{\Lambda_{b}}\bar{\sigma}^{2}-m^{2}_{u})}{m^{2}_{\Lambda_{b}}\bar{\sigma}}f^{(2)}_{\Lambda_b}\overline{\psi_{2}}+\frac{8(m_{u}+\sigma m_{\Lambda_{b}})}{m_{\Lambda_{b}}}f^{(1)}_{\Lambda_b}\overline{\psi^{\sigma}_{3}}\big]\ ,\nonumber\\
C^{V}_{6,1}&&=\big[\frac{12}{m^{2}_{\Lambda_{b}}\bar{\sigma}}f^{(2)}_{\Lambda_b}\overline{\psi_{2}}\big]\ ,\nonumber\\
C^{V}_{7,2}&&=\big[16\sigma f^{(1)}_{\Lambda_b}\overline{\psi^{\sigma}_{3}}
        +24\sigma^{2} f^{(2)}_{\Lambda_b}(\overline{\psi_{4}}-\overline{\psi_{2}})+\frac{24\sigma(q^{2}-m^{2}_{\Lambda_{b}}\bar{\sigma}^{2}-m^{2}_{u})}{m^{2}_{\Lambda_{b}}\bar{\sigma}}f^{(2)}_{\Lambda_b}\overline{\psi_{2}}\big]\ ,\nonumber\\
C^{V}_{7,1}&&=-\big[\frac{24}{m^{2}_{\Lambda_{b}}\bar{\sigma}}f^{(2)}_{\Lambda_b}\overline{\psi_{2}}\big]\ ,       
\end{eqnarray}
where the functions $\overline{\psi}(\omega,u)$ are defined as:
\begin{eqnarray}
	\overline{\psi}(\omega, u)=\int^{\omega}_0 d\tau \ \tau\psi(\tau,u)\ ,
\end{eqnarray}
originating from the partial integral in the variable $\omega$ in Eq.(\ref{EQ:invaramp})
\item For the axlai-vector current $J^{A}_{\mu}=\bar{u}\gamma_{\mu}\gamma_{5}b$:
\begin{eqnarray}
C^{\rm A}_{1,n}&&=C^{\rm A}_{2,n}(n=1-2)=0\ , \ \ C^{A}_{3,1}=C^{A}_{4,1}=C^{A}_{8,1}=0\ ,\nonumber\\
C^{\rm A}_{3,2}&&= -\big[24\sigma\bar{\sigma}f^{(2)}_{\Lambda_b}(\overline{\psi_{4}}-\overline{\psi_{2}})+\frac{24(q^{2}-m^{2}_{\Lambda_{b}}\bar{\sigma}^{2}-m^{2}_{u})}{m^{2}_{\Lambda_{b}}}f^{(2)}_{\Lambda_b}\overline{\psi_{2}}+\frac{16m_{u}}{m_{\Lambda_{b}}}f^{(1)}_{\Lambda_b}\overline{\psi^{\sigma}_{3}}\big]\ ,\nonumber \\
C^{A}_{4,2}&&=C^{A}_{8,2}=\big[\frac{16\sigma}{m_{\Lambda_{b}}}f^{(1)}_{\Lambda_b}\overline{\psi^{\sigma}_{3}}\big]\ ,\nonumber \\
C^{A}_{5,2}&&=\big[12\sigma (m_{u}+\bar{\sigma}m_{\Lambda_{b}})f^{(2)}_{\Lambda_b}(\overline{\psi_{4}}-\overline{\psi_{2}})+\frac{8(m^{2}_{u}-\sigma q^{2}+m_{u}m_{\Lambda_{b}}\bar{\sigma})}{m_{\Lambda_{b}}\bar{\sigma}}f^{(1)}_{\Lambda_b}\overline{\psi^{\sigma}_{3}}\nonumber\\
&&\ \ \ +\frac{12(q^{2}-m^{2}_{\Lambda_{b}}\bar{\sigma}^{2}-m^{2}_{u})(m_{u}+\bar{\sigma}m_{\Lambda_{b}})}{m^{2}_{\Lambda_{b}}\bar{\sigma}}f^{(2)}_{\Lambda_b}\overline{\psi_{2}} \big]\ ,\nonumber \\
C^{A}_{5,1}&&=\big[-\frac{12m_{u}}{m^{2}_{\Lambda_{b}}\bar{\sigma}}f^{(2)}_{\Lambda_b}\overline{\psi_{2}}+\frac{8\sigma}{m_{\Lambda_{b}}\bar{\sigma}}f^{(1)}_{\Lambda_b}\overline{\psi^{\sigma}_{3}}\big]\ ,\nonumber \\
C^{A}_{6,2}&&=-\big[12\sigma f^{(2)}_{\Lambda_b}(\overline{\psi_{4}}-\overline{\psi_{2}})+\frac{12(q^{2}-m^{2}_{\Lambda_{b}}\bar{\sigma}^{2}-m^{2}_{u})}{m^{2}_{\Lambda_{b}}\bar{\sigma}}f^{(2)}_{\Lambda_b}\overline{\psi_{2}}+\frac{8(m_{u}-\sigma m_{\Lambda_{b}})}{m_{\Lambda_{b}}}f^{(1)}_{\Lambda_b}\overline{\psi^{\sigma}_{3}}\big]\ ,\nonumber\\
C^{A}_{6,1}&&=\big[\frac{12}{m^{2}_{\Lambda_{b}}\bar{\sigma}}f^{(2)}_{\Lambda_b}\overline{\psi_{2}}\big]\ ,\nonumber\\
C^{A}_{7,2}&&=\big[-16\sigma f^{(1)}_{\Lambda_b}\overline{\psi^{\sigma}_{3}}
        +24\sigma^{2} f^{(2)}_{\Lambda_b}(\overline{\psi_{4}}-\overline{\psi_{2}})+\frac{24\sigma(q^{2}-m^{2}_{\Lambda_{b}}\bar{\sigma}^{2}-m^{2}_{u})}{m^{2}_{\Lambda_{b}}\bar{\sigma}}f^{(2)}_{\Lambda_b}\overline{\psi_{2}}\big]\ ,\nonumber\\
C^{A}_{7,1}&&=-\big[\frac{24}{m^{2}_{\Lambda_{b}}\bar{\sigma}}f^{(2)}_{\Lambda_b}\overline{\psi_{2}}\big] \ ,      
\end{eqnarray}
\end{itemize}

\end{document}